\documentclass{article}
\usepackage{spconf,amsmath,graphicx,amssymb}
\usepackage{algorithm}
\usepackage{algorithmicx, algpseudocode }
\usepackage{multirow}
\usepackage{subcaption}

\title{Dynamic Polygon Clouds: Representation and Compression for VR/AR \\ {\small\em Microsoft Research Technical Report MSR-TR-2016-59} \\ {\small\em (Draft as of January 4, 2017 --- see arxiv.org for updates)}}
\name{
Philip A. Chou$^1$\thanks{P. A. Chou is with Microsoft Research, Redmond, WA, USA, e-mail: pachou@ieee.org.},
Eduardo Pavez $^2$\thanks{E. Pavez is with the Department of Electrical Engineering, University of Southern California, Los Angeles, CA, USA, e-mail: pavezcar@usc.edu},
Ricardo L. de Queiroz$^3$\thanks{R. L. de Queiroz is with the Computer Science Department at Universidade de Brasilia, Brasilia, Brazil, e-mail: queiroz@ieee.org.},
and\ Antonio Ortega$^2$\thanks{Antonio Ortega is with the Department of Electrical Engineering at the University of Southern California, Los Angeles, CA, USA, e-mail: antonio.ortega@sipi.usc.edu}
}
\address{
$^1$Microsoft Research, Redmond, WA, USA \\
$^2$University of Southern California, Los Angeles, CA, USA \\
$^3$Universidade de Brasilia, Brasilia, Brazil
}
\begin{document}
%
\maketitle
\begin{abstract}
We introduce the {\em polygon cloud}, also known as a polygon set or {\em soup}, as a compressible representation of 3D geometry (including its attributes, such as color texture) intermediate between polygonal meshes and point clouds.  Dynamic or time-varying polygon clouds, like dynamic polygonal meshes and dynamic point clouds, can take advantage of temporal redundancy for compression, if certain challenges are addressed.  In this paper, we propose methods for compressing both static and dynamic polygon clouds, specifically triangle clouds.  We compare triangle clouds to both triangle meshes and point clouds in terms of compression, for live captured dynamic colored geometry.  We find that triangle clouds can be compressed nearly as well as triangle meshes, while being far more robust to noise and other structures typically found in live captures, which violate the assumption of a smooth surface manifold, such as lines, points, and ragged boundaries.  We also find that triangle clouds can be used to compress point clouds with significantly better performance than previously demonstrated point cloud compression methods.  In particular, for intra-frame coding of geometry, our method improves upon octree-based intra-frame coding by a factor of 5-10 in bit rate.  Inter-frame coding improves this by another factor of 2-5.  Overall, our dynamic triangle cloud compression improves over the previous state-of-the-art in dynamic point cloud compression by 33\% or more.
\end{abstract}
\section{Introduction}
\label{sec:intro}

With the advent of virtual and augmented reality comes the birth of a new medium: live captured 3D content that can be experienced from any point of view.  Such content ranges from static scans of compact 3D objects, to dynamic captures of non-rigid objects such as people, to captures of rooms including furniture, public spaces swarming with people, and whole cities in motion.  For such content to be captured at one place and delivered to another for consumption by a virtual or augmented reality device (or by more conventional means), the content needs to be represented and compressed for transmission or storage.  Applications include gaming, tele-immersive communication, free navigation of highly produced entertainment as well as live events,  historical artifact and site preservation, acquisition for special effects, and so forth.  This paper presents a novel means of representing and compressing the visual part of such content.

Until this point, two of the more promising approaches to representing both static and time-varying 3D scenes have been polygonal meshes and point clouds, along with their associated color information.  However, both approaches have drawbacks.  Polygonal meshes represent surfaces very well, but they are not robust to noise and other structures typically found in live captures, such as lines, points, and ragged boundaries that violate the assumptions of a smooth surface manifold.  Point clouds, on the other hand, have a hard time modeling surfaces as compactly as meshes.

We propose a hybrid between polygonal meshes and point clouds: polygon clouds.  Polygon clouds are sets of polygons, often called a polygon soup.  The polygons in a polygon cloud are not required to represent a coherent surface.  Like the points in a point cloud, the polygons in a polygon cloud can represent noisy, real-world geometry captures without any assumption of a smooth 2D manifold.  In fact, any polygon in a polygon cloud can be collapsed into a point or line as a special case.  The polygons may also overlap.  On the other hand, the polygons in the cloud can also be stitched together into a watertight mesh if desired to represent a smooth surface.  Thus polygon clouds generalize both point clouds and polygonal meshes.

For concreteness we focus on triangles instead of arbitrary polygons, and we develop an encoder and decoder for sequences of triangle clouds.  We assume a simple group of frames (GOF) model, where each group of frames begins with an Intra (I) frame, also called a reference frame or a key frame, which is followed by a sequence of Predicted (P) frames, also called inter frames.  The triangles are assumed to be consistent across frames.  That is, the triangles' vertices are assumed to be tracked from one frame to the next.  The trajectories of the vertices are not constrained.  Thus the triangles may change from frame to frame in location, orientation, and proportion.  For geometry encoding, redundancy in the vertex trajectories is removed by a spatial othogonal transform followed by temporal prediction, allowing low latency.  For color encoding, the triangles in each frame are projected back to the coordinate system of the reference frame.  In the reference frame we voxelize the triangles in order to ensure that their color textures are sampled uniformly in space regardless of the sizes of the triangles, and in order to construct a common vector space in which to describe the color textures and their evolution from frame to frame.  Redundancy of the color vectors is removed by a spatial orthogonal transform followed by temporal prediction, similar to redundancy removal for geometry.  Uniform scalar quantization and entropy coding matched to the spatial transform are employed for both color and geometry.

We compare triangle clouds to both triangle meshes and point clouds in terms of compression, for live captured dynamic colored geometry.  We find that triangle clouds can be compressed nearly as well as triangle meshes, while being far more flexible in representing live captured content.  We also find that triangle clouds can be used to compress point clouds with significantly better performance than previously demonstrated point cloud compression methods.

The organization of the paper is as follows.  Following a summary of related work in Section~\ref{sec:related}, preliminary material is presented in Section~\ref{sec:preliminaries}.  Components of our compression system are presented in Section~\ref{sec:components}, while the core of our system is presented in Section~\ref{sec:system}. Experimental results are presented in Section~\ref{sec:experiments}.  The conclusion is in Section~\ref{sec:discussion}.

\section{Related work}
\label{sec:related}

\subsection{Mesh compression}

3D mesh compression has a rich history, particularly from the 1990s forward. Overviews may be found in \cite{AlliezG05,Peng_2005,MagloLDH13}.  Fundamental is the need to code mesh topology, or connectivity, such as in \cite{Rossignac_1999,MamouZP09}.  Beyond coding connectivity, coding the geometry, i.e., the positions of the vertices, is also fundamental.  Many approaches have been taken, but one significant and practical approach to geometry coding is based on ``geometry images'' \cite{GuGH02} and their temporal extension, ``geometry videos'' \cite{BricenoSMGH03}.  In these approaches, the mesh is partitioned into patches, the patches are projected onto a 2D plane as {\em charts}, non-overlapping charts are laid out in a rectangular {\em atlas}, and the atlas is compressed using a standard image or video coder, compressing both the geometry and the texture (i.e., color) data.  For dynamic geometry, the meshes are assumed to be temporally consistent (i.e., connectivity is constant frame-to-frame) and the patches are likewise temporally consistent.  Geometry videos have been used for representing and compressing free-viewpoint video of human actors \cite{ColletCSGECHKS15}.  Other key papers on mesh compression of human actors in the context of tele-immersion include \cite{MekuriaSIBC14,DoumanoglouAZD14}.

\subsection{Motion estimation}

A critical part of dynamic mesh compression is the ability to track points over time.  If a mesh is defined for a keyframe, and the vertices are tracked over subsequent frames, then the mesh becomes a temporally consistent dynamic mesh. There is a huge body of literature in the 3D tracking, 3D motion estimation or scene flow, 3D interest point detection and matching, 3D correspondence, non-rigid registration, and the like.  We are particularly influenced by \cite{NewcombeFS15,DouTFFI15,dou2016fusion4d}, all of which produce in real time, given data from one or more RGBD sensors for every frame $t$, a parameterized mapping $f_{\theta_t}:{\mathbb R}^3\rightarrow {\mathbb R}^3$ that maps points in frame $t$ to points in frame $t+1$. Though corrections may need to be made at each frame, chaining the mappings together over time yields trajectories for any given set of points.  Compressing these trajectories is similar to compressing motion capture (mocap) trajectories, which has been well studied. \cite{HouCMH15} is a recent example with many references. Compression typically involves an intra-frame transform to remove spatial redundancy and either temporal prediction (if low latency is required) or a temporal transform (if the entire clip or group of frames is available) to remove temporal redundancy, as in \cite{HouCMH16}.

\subsection{Graph signal processing}

Graph Signal Processing (GSP) has emerged as an extension of the theory of linear shift invariant signal processing to the processing of signals on discrete graphs, where the shift operator is taken to be the adjacency matrix of the graph, or alternatively the Laplacian matrix of the graph \cite{SandryhailaM13,Shuman2013}.  GSP was extended to critically sampled perfect reconstuction wavelet filter banks in \cite{NarangO12,NarangO13}.  These constructions were used for dynamic mesh compression in \cite{nguyen_2014,anis_2016}.

\subsection{Point cloud compression using octrees}

Sparse Voxel Octrees (SVOs) were developed in the 1980s to represent the geometry of three-dimensional objects \cite{JackinsT80,meagher_octtree}.  Recently SVOs have been shown to have highly efficient implementations suitable for encoding at video frame rates \cite{Loop_2013}.  In the guise of occupancy grids, they have also had significant use in robotics \cite{Moravec88,Elfes89,PathakBPS07}.  Octrees were first used for point cloud compression in \cite{schnabel_2006}.  They were further developed for progressive point cloud coding, including color attribute compression, in \cite{Huang2008}.  Octrees were extended to coding of dynamic point clouds (i.e., point cloud sequences) in \cite{Kammerl2012}.  The focus of \cite{Kammerl2012} was geometry coding; their color attribute coding remained rudimentary.  Their method of inter-frame geometry coding was to take the exclusive-OR (XOR) between frames and code the XOR using an octree.  Their method was implemented in the Point Cloud Library \cite{Rusu_3dis}.

\subsection{Color attribute compression for static point clouds}

To better compress the color attributes in {\em static} voxelized point clouds, Zhang, Flor\^encio, and Loop used transform coding based on the Graph Fourier Transform (GFT), recently developed in the theory of Graph Signal Processing \cite{ZhangFL14}.  While transform coding based on the GFT has good compression performance, it requires eigen-decompositions for each coded block, and hence may not be computationally attractive.  To improve the computational efficiency, while not sacrificing compression performance, Queiroz and Chou developed an orthogonal Region-Adaptive Hierarchical Transform (RAHT) along with an entropy coder \cite{queiroz_raht}.  RAHT is essentially a Haar transform with the coefficients appropriately weighted to take the non-uniform shape of the domain (or region) into account.  As its structure matches the Sparse Voxel Octree, it is extremely fast to compute.  Other approaches to non-uniform regions include the shape-adaptive DCT \cite{CohenTV16} and color palette coding \cite{DadoKBTE16}.  Further approaches based on non-uniform sampling of an underlying stationary process can be found in \cite{QueirozC16-klt}, which uses the KLT matched to the sample, and in \cite{HouCHC17}, which uses sparse representation and orthogonal matching pursuit.

\subsection{Dynamic point cloud compression}

Thanou, Chou, and Frossard \cite{thanou_icip_2015,ThanouCF16} were the first to deal fully with {\em dynamic} voxelized points clouds, by finding matches between points in adjacent frames, warping the previous frame to the current frame, predicting the color attributes of the current frame from the quantized colors of the previous frame, and coding the residual using the GFT-based method of \cite{ZhangFL14}.  Thanou et al.\ used the XOR-based method of Kammerl et al.\ \cite{Kammerl2012} for inter-frame geometry compression.  However, the method of \cite{Kammerl2012} proved to be inefficient, in a rate-distortion sense, for anything except slowly moving subjects, for two reasons.  First, the method ``predicts'' the current frame from the previous frame, without any motion compensation.  Second, the method codes the geometry losslessly, and so has no ability to perform a rate-distortion trade-off.  To address these shortcomings, Queiroz and Chou \cite{QueirozC16-simple} used block-based motion compensation and rate-distortion optimization to select between coding modes (intra or motion-compensated coding) for each block.  Further, they applied RAHT to coding the color attributes (in intra-frame mode), color prediction residuals (in inter-frame mode), and the motion vectors (in inter-frame mode).  They also used in-loop deblocking filters.  Mekuria et al.\ \cite{mekuria_2016} independently proposed block-based motion compensation for dynamic point cloud sequences.  Although they did not use rate-distortion optimization, they used affine transformations for each motion-compensated block, rather than just translations.  Unfortunately, it appears that block-based motion compensation of dynamic point cloud geometry tends to produce gaps between blocks, which are perceptually more damaging than indicated by objective metrics such as the Haussdorf-based metrics commonly used in geometry compression \cite{MekuriaLTC16}.

\subsection{Key learnings}

Some of the key learnings from the previous work, taken as a whole, are that
\begin{itemize}
\item Point clouds are preferable to meshes for resilience to noise and non-manifold signals measured in real world signals, especially for real time capture where the computational cost of heavy duty pre-processing (e.g., surface reconstruction, topological denoising, charting) can be prohibitive.
\item For geometry coding in static scenes, point clouds appear to be more compressible than meshes, even though the performance of point cloud geometry coding seems to be limited by the lossless nature of the current octree methods.  In addition, octree processing for geometry coding is extremely fast.
\item For color attribute coding in static scenes, both point clouds and meshes appear to be well compressible.  If charting is possible, compressing the color as an image may win out due to the maturity of image compression algorithms today.  However, direct octree processing for color attribute coding is extremely fast, as it is for geometry coding.
\item For both geometry and color attribute coding in dynamic scenes (or inter-frame coding), temporally consistent dynamic meshes are highly compressible.  However, finding a temporally consistent mesh can be challenging from a topological point of view as well as from a computational point of view.
\end{itemize}
In our work, we aim to achieve the high compression efficiency possible with intra-frame point cloud compression and inter-frame dynamic mesh compression, while simultaneously achieving the high computational efficiency possible with octree-based processing, as well as its robustness to real-world noise and non-manifold data.

\section{Preliminaries}
\label{sec:preliminaries}

\subsection{Notation}
Notation is given in Table~\ref{tab:notation}.
\begin{table}[!h]
\begin{tabular}{|c|c|} \hline
symbol & description \\ \hline 
$[N]$ & set of integers $\lbrace 1,2,\cdots, N \rbrace$ \\ \hline
$t$ & time or frame index \\ \hline
$v_i$ or $v_i^{(t)}$ & 3D point with coordinates $x_i,y_i,z_i$ \\ \hline
$f_m$ or $f_m^{(t)}$ &  face with vertex indices $i_m,j_m,k_m $\\ \hline
$c_n$ or $c_n^{(t)}$ & color with components $Y_n,U_n,V_n$ \\ \hline
$a_i$ or $a_i^{(t)}$ & generic attribute vector $a_{i1},\ldots,a_{in}$ \\ \hline
$\mathcal{V}$ or $\mathcal{V}^{(t)}$ & set of $N_p$ points $\lbrace v_1,\ldots,v_{N_p} \rbrace$ \\ \hline
$\mathcal{F}$ or $\mathcal{F}^{(t)}$ & set of $N_f$ faces $\lbrace f_1,\ldots,f_{N_f} \rbrace$ \\ \hline
$\mathcal{C}$ or $\mathcal{C}^{(t)}$ & set of $N_c$ colors $\lbrace c_1,\ldots,c_{N_c} \rbrace$ \\ \hline
$\mathcal{A}$ or $\mathcal{A}^{(t)}$ & set of $N_a$ attribute vectors $\lbrace a_1,\ldots,a_{N_a} \rbrace$ \\ \hline
$\mathcal{T}$ or $\mathcal{T}^{(t)}$ & triangle cloud $(\mathcal{V},\mathcal{F},\mathcal{C})$ or $(\mathcal{V},\mathcal{F},\mathcal{A})$ \\ \hline
$\mathcal{P}$ or $\mathcal{P}^{(t)}$ & point cloud $(\mathcal{V},\mathcal{C})$ or $(\mathcal{V},\mathcal{A})$ \\ \hline
$\mathbf{V}$ or $\mathbf{V}^{(t)}$ & $N_p \times 3$ matrix with $i$-th row $[x_i,y_i,z_i]$ \\ \hline
$\mathbf{F}$ or $\mathbf{F}^{(t)}$ & $N_f \times 3$ matrix with $m$-th row $[i_m,j_m,k_m]$ \\ \hline
$\mathbf{C}$ or $\mathbf{C}^{(t)}$ & $N_c \times 3$ matrix with $n$-th row $[Y_n,U_n,V_n]$ \\ \hline
$\mathbf{A}$ & list (i.e., matrix) of attributes \\ \hline
$\mathbf{TA}$ & list of transformed attributes \\ \hline
$\mathbf{M},\!\mathbf{M}_v,\!\mathbf{M}_1$ & lists of Morton codes \\ \hline
{\small $\mathbf{W},\!\!\mathbf{W}_v,\!\!\mathbf{W}_{rv}$} & lists of weights \\ \hline
$\mathbf{I},\mathbf{I}_v,\mathbf{I}_{rv}$ & lists of indices \\ \hline
$\hat{\mathbf{V}},\!\hat{\mathbf{C}},\!\hat{\mathbf{A}},\!\ldots$ & lists of quantized or reproduced quantities \\ \hline
$\hat{\mathbf{V}}_v$ or $\hat{\mathbf{V}}_v^{(t)}$ & list of voxelized vertices \\ \hline
$\mathbf{V}_r$ & list of refined vertices \\ \hline
$\hat{\mathbf{V}}_{rv}$\! or\! $\hat{\mathbf{V}}_{rv}^{(t)}$ & list of voxelized refined vertices \\ \hline
$\mathbf{C}_r=\mathbf{C}$ & list of colors of refined vertices \\ \hline
$\mathbf{C}_{rv}$\! or\! $\mathbf{C}_{rv}^{(t)}$ & list of colors of voxelized refined vertices \\ \hline
$J$ & octree depth \\ \hline
$U$ & upsampling factor \\ \hline
$\Delta_{motion}$ & motion quantization stepsize \\ \hline
$\Delta_{color,intra}$ & intra-frame color quantization stepsize \\ \hline
$\Delta_{color,inter}$ & inter-frame color quantization stepsize \\ \hline
\end{tabular}
\caption{Notation }
\label{tab:notation}
\end{table}
\subsection{Dynamic triangle clouds}
A dynamic triangle cloud is a numerical representation of a time changing 3D scene or object. We denote it by a sequence $\lbrace \mathcal{T}^{(t)} \rbrace$ where $\mathcal{T}^{(t)}$ is a triangle cloud at time $t$. Each individual frame $\mathcal{T}^{(t)}$ has geometry (shape and position) and color information.

The geometry information consists of a list of vertices $\mathcal{V}^{(t)}=\lbrace v_i^{(t)} : i=1,\cdots, N_p\rbrace $, where each vertex $v_i^{(t)}=[x_i^{(t)},y_i^{(t)},z_i^{(t)}]$ is a point in 3D, and a list of triangles (or faces) $\mathcal{F}^{(t)}=\lbrace f_m^{(t)}: m=1,\cdots,N_f \rbrace$, where each face $f_m^{(t)}=[i_m^{(t)},j_m^{(t)},k_m^{(t)}]$ is a vector of indices of vertices from $\mathcal{V}^{(t)}$.  We denote by $\mathbf{V}^{(t)}$ the $N_p \times 3$ matrix whose $i$-th row is the point $v_i^{(t)}$, and similarly we denote by $\mathbf{F}^{(t)}$ the $N_f \times 3$ matrix whose $m$-th row is the triangle $f_m^{(t)}$.  The triangles in a triangle cloud do not have to be adjacent or form a  mesh, and they can overlap. Two or more vertices of a triangle may have the same coordinates, thus collapsing into a line or point. 
 
\begin{figure*} 
\begin{subfigure}[b]{0.3\textwidth}
\hspace{-3.5cm}
\includegraphics[scale=0.48]{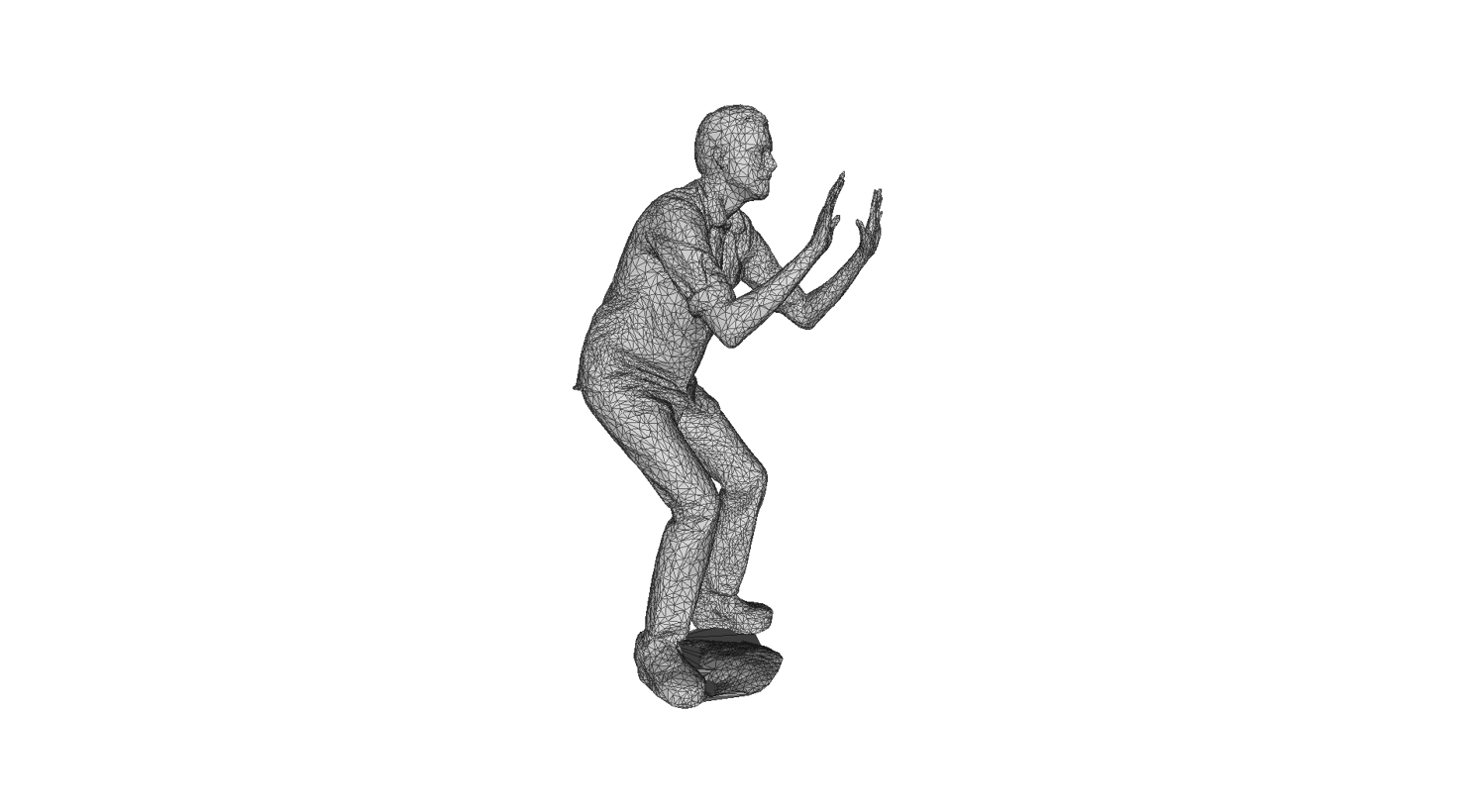}
\caption{\emph{Man} mesh.}
\label{fig_man_mesh}
\end{subfigure}
\begin{subfigure}[b]{0.48\textwidth}
\includegraphics[scale=0.31]{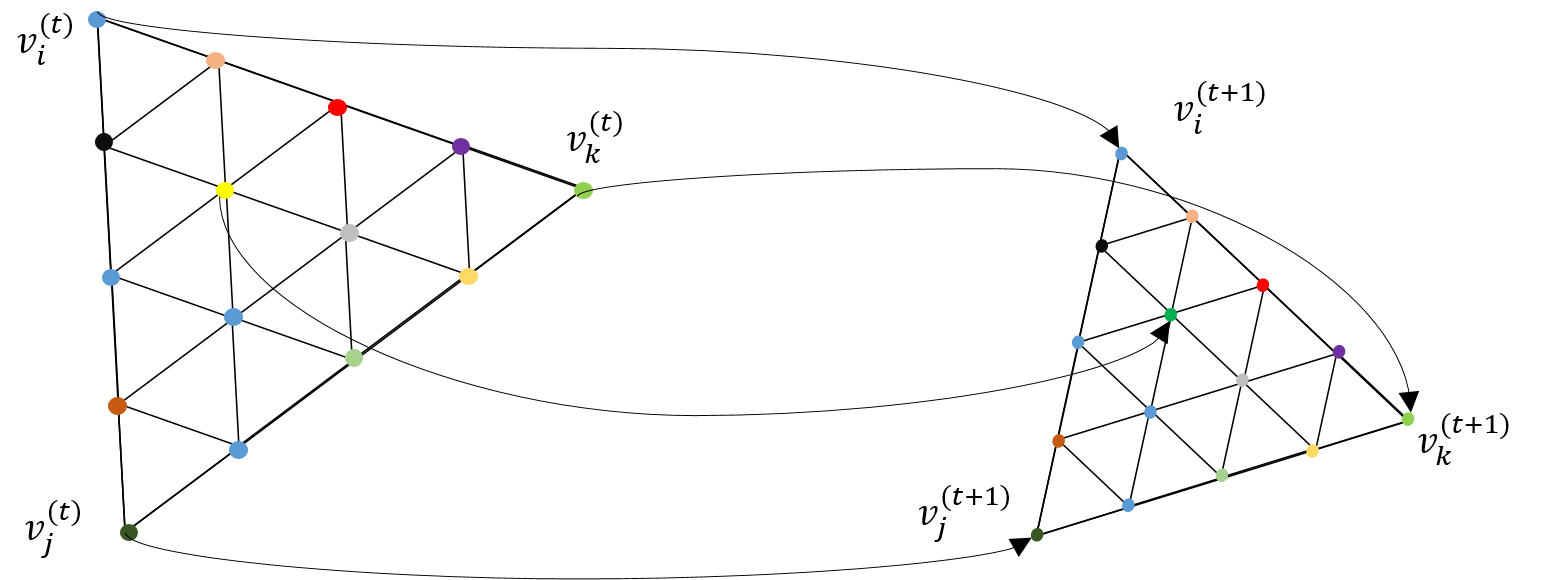}
\vspace{0.5cm}
\caption{Correspondences between two consecutive frames.}
\label{fig_correspondences}
\end{subfigure}
\caption{Triangle cloud geometry information.}
\end{figure*}

The color information consists of a list of colors $\mathcal{C}^{(t)}=\lbrace c_n^{(t)} : n=1,\cdots, N_c\rbrace $, where each color $c_n^{(t)}=[Y_n^{(t)},U_n^{(t)}$, $V_n^{(t)}]$ is a vector in YUV space (or other convenient color space).  We denote by $\mathbf{C}^{(t)}$ the $N_c \times 3$ matrix whose $n$-th row is the color $c_n^{(t)}$.  The list of colors represents the colors across the surfaces of the triangles.  To be specific, $c_n^{(t)}$ is the color of a ``refined'' vertex $v_r^{(t)}(n)$, where the refined vertices are obtained by uniformly subdividing each triangle in $\mathcal{F}^{(t)}$ by upsampling factor $U$, as shown in Figure~\ref{fig_correspondences} for $U=4$.  We denote by $\mathbf{V}_r^{(t)}$ the $N_c \times 3$ matrix whose $n$th row is the refined vertex $v_r^{(t)}(n)$.  $\mathbf{V}_r^{(t)}$ can be computed from $\mathcal{V}^{(t)}$ and $\mathcal{F}^{(t)}$, so we do not need to encode it, but we will use it to compress the color information.  Note that $N_c = N_f(U+1)(U+2)/2$.  The upsampling factor $U$ should be high enough so that it does not limit the color spatial resolution obtainable by the color cameras.  In our experiments, we set $U=10$ or higher.  Setting $U$ higher does not typically affect the bit rate significantly, though it does affect memory and computation in the encoder and decoder.

Thus frame $t$ can be represented by the triple $\mathbf{V}^{(t)}$, $\mathbf{F}^{(t)}$, $\mathbf{C}^{(t)}$. We use a Group of Frames (GOF) model, in which the sequence is partitioned into GOFs.  The GOFs are processed independently.  Without loss of generality, we label the frames in a GOF $t=1\ldots,N$.  There are two types of frames: reference and predicted.  In each GOF, the first frame ($t=1$) is a reference frame and all other frames ($t=2,\ldots,N$) are predicted.  Within a GOF, all frames must have the same number of vertices, triangles, and colors:  $\forall t\in [N]$, $\mathbf{V}^{(t)} \in \mathbb{R}^{N_p\times 3}$, $\mathbf{F}^{(t)} \in [N_p]^{N_f \times 3}$ and  $\mathbf{C}^{(t)} \in \mathbb{R}^{N_c \times 3}$.  The triangles are assumed to be consistent across frames so that there is a correspondence between colors and vertices within the GOF.  In Figure \ref{fig_correspondences} we show an example of the correspondences between two consecutive frames in a GOF.  Across GOFs, the GOFs may have a different numbers of frames, vertices, triangles, and colors.

In the following two subsections, we outline how to obtain a triangle cloud from an existing point cloud or an existing triangular mesh.

\subsubsection{Converting a dynamic point cloud to a dynamic triangle cloud}
A dynamic point cloud is a sequence of point clouds $\lbrace \mathcal{P}^{(t)}\rbrace $, where each $\mathcal{P}^{(t)}$ is a list of  $[x,y,z]$ coordinates each with an attribute attached to it, like color. To produce a triangle cloud, we need a way to fit a point cloud to a set of triangles in such a way that we produce  GOFs with consistent triangles. One way of doing that is the following.
\begin{enumerate}
\item Decide if frame in $\mathcal{P}^{(t)}$ is reference or predicted.
\item If reference frame:
\begin{enumerate}
\item Fit triangles to point cloud to obtain $\mathbf{V}^{(1)},\mathbf{F}^{(1)}$, where $\mathbf{V}^{(1)}$ is a list of vertices and $\mathbf{F}^{(1)}$  is a list of triangles. 
\item Subdivide each triangle, and project each vertex of the subdivision to the closest point in the cloud to obtain $\mathbf{C}^{(1)}$.
\end{enumerate}
\item If predicted frame:
\begin{enumerate}
\item Deform triangle cloud of previous reference frame to fit point cloud to obtain $\mathbf{V}^{(t)}$, such that the $i$th point $v_i^{(t)}$ in $\mathbf{V}^{(t)}$ corresponds to the $i$th point $v_i^{(1)}$ in $\mathbf{V}^{(1)}$.
\item Subdivide each triangle, and project each vertex of the subdivision to the closest point in the cloud to obtain $\mathbf{C}^{(t)}$.
\item Go to step 1.
\end{enumerate}
\end{enumerate}
This process will introduce geometric distortion and a change in the number of points. All points will be forced to lie in a uniform grid on the surface of a triangle. The triangle fitting can be done using triangular mesh fitting and tracking techniques such as in \cite{NewcombeFS15,DouTFFI15,dou2016fusion4d}.
\subsubsection{Converting a dynamic triangular mesh to a dynamic triangle cloud}
The geometry of a  triangular mesh is represented by a list of key points or vertices and their connectivity, given by an array of 3D coordinates $\mathbf{V}$ and faces $\mathbf{F}$. The triangles are constrained to form a smooth surface without holes.  For color, the mesh representation typically includes an array of 2D texture coordinates $\mathbf{T}\in\mathbb{R}^{N_p\times 2}$ and a texture image.  The color at any point on a face can be retrieved (for rendering) by interpolating the texture coordinates at that point on the face and sampling the image at the interpolated coordinates. The sequence of triangular meshes is assumed to be temporally consistent, meaning that within a GOF, the meshes of the predicted frames are deformations of the reference frame.  The sizes and positions of the triangles may change but the deformed mesh still represents a smooth surface. The sequence of key points $\mathbf{V}^{(t)}$ thus can be traced from frame to frame and the faces are all the same. To convert the color information into the dynamic triangle cloud format, for each frame and each triangle, the mesh sub-division function can be applied to obtain texture coordinates of refined triangles. Then the texture image can be sampled and a color matrix $\mathbf{C}$ can be formed for each frame.

\subsection{Compression  system overview}

In this section we provide an overview of our system for compressing dynamic triangle clouds.  We compress consecutive GOFs sequentially and independently, so we focus on the system for compressing an individual GOF  $(\mathbf{V}^{(t)},\mathbf{F}^{(t)},\mathbf{C}^{(t)})$ for $t \in [N]$.

For the reference frame, we voxelize the vertices $\mathbf{V}^{(1)}$, and then encode the voxelized vertices $\mathbf{V}_v^{(1)}$ using octree encoding.  We encode the connectivity $\mathbf{F}^{(1)}$ with a lossless entropy coder.  (We could use method such as EdgeBreaker or TFAN \cite{Rossignac_1999,MamouZP09}, but for simplicity for this small amount of data in our experiments we use the lossless universal encoder \emph{gzip}.)   We code the connectivity only once per GOF (i.e., for the reference frame), since the connectivity is consistent across the GOF, i.e., $\mathbf{F}^{(t)}=\mathbf{F}^{(1)}$ for $t\in[N]$.  We voxelize the colors $\mathbf{C}^{(1)}$, and encode the voxelized colors $\mathbf{C}_{rv}^{(1)}$ using a transform coding method that combines the region adaptive hierarchical transform (RAHT) \cite{queiroz_raht}, uniform scalar quantization, and adaptive Run-Length Golomb-Rice (RLGR) entropy coding \cite{malvar_rlgr}.  At the cost of additional complexity, the RAHT transform could be replaced by transforms with higher performance \cite{QueirozC16-klt,HouCHC17}.

For predicted frames, we compute prediction residuals from the previously decoded frame. Specifically, for each predicted frame $t>1$ we compute a motion residual $\Delta \mathbf{V}_v^{(t)}=\mathbf{V}_v^{(t)}-\hat{\mathbf{V}}_v^{(t-1)}$ and a color residual $\Delta \mathbf{C}_{rv}^{(t)}=\mathbf{C}_{rv}^{(t)}-\hat{\mathbf{C}}_{rv}^{(t-1)}$, where we have denoted with a \emph{hat} a quantity that has been compressed and decompressed. These residuals are encoded using again RAHT followed by uniform scalar quantization and entropy coding.

It is important to note that we do not directly compress the list of vertices $\mathbf{V}^{(t)}$ or the the list of colors $\mathbf{C}^{(t)}$ (or their prediction residuals).  Rather, we voxelize them first {\em with respect to their corresponding vertices in the reference frame}, and then compress them.  This ensures that 1) if two or more vertices or colors fall into the same voxel, they receive the same representation and hence are encoded only once, and 2) the colors (on the set of refined vertices) are resampled uniformly in space regardless of the density or shapes of triangles.

In the next section, we detail the basic elements of the system: refinement, voxelization, octrees, and transform coding.  In the section after that, we detail how these basic elements are put together to encode and decode a sequence of triangle clouds.




\section{Refinement, voxelization, octrees, and transform coding}
\label{sec:components}

\subsection{Refinement}

Given a list of faces $\mathbf{F}$, its corresponding list of vertices $\mathbf{V}$, and upsampling factor $U$, a list of ``refined'' vertices $\mathbf{V}_r$ can be produced using Algorithm~\ref{alg:refine}.  Step~1 (in Matlab notation) assembles three equal-length lists of vertices (each as an $N_f\times3$ matrix), containing the three vertices of every face.  Step~5 appends a linear combinations of the faces' vertices to a growing list of refined vertices.

\begin{algorithm}
\caption{Refinement ({\em refine})}
\label{alg:refine}
\begin{algorithmic}[1]
\Require $\mathbf{V}$, $\mathbf{F}$, $U$
\State $\mathbf{V}_i=\mathbf{V}(\mathbf{F}(:,i),:)$, $i=1,2,3$ // $i$th vertex of all faces
\State Initialize $\mathbf{V}_r=$ empty list
\For{$i=0$ to $U$}
	\For{$j=0$ to $U-i$}
    	\State $\mathbf{V}_r=[\mathbf{V}_r;\mathbf{V}_1+(\mathbf{V}_2-\mathbf{V}_1)i/U+(\mathbf{V}_3-\mathbf{V}_1)j/U]$
    \EndFor
\EndFor
\Ensure $\mathbf{V}_r$
\end{algorithmic}
\end{algorithm}

We assume that the list of colors $\mathbf{C}$ is in 1-1 correspondence with the list of refined vertices $\mathbf{V}_r$.  Indeed, to obtain the colors $\mathbf{C}$ from a textured mesh, the 2D texture coordinates $\mathbf{T}$ can be linearly interpolated in the same manner as the 3D position coordinates $\mathbf{V}$ to obtain ``refined'' texture coordinates $\mathbf{T}_r$ which may then be used to lookup appropriate color $\mathbf{C}_r=\mathbf{C}$ in the texture map.

\subsection{Morton codes and voxelization}

A \emph{voxel} is a volumetric element used to represent the attributes of an object in 3D over a small region of space.  Analogous to 2D pixels, 3D voxels are defined on a uniform grid.  We assume the geometric data live in the unit cube $[0,1)^3$, and we uniformly partition the cube into voxels of size $2^{-J} \times 2^{-J} \times 2^{-J}$.

Now consider a list of points $\mathbf{V}=[v_i]$ and an equal-length list of attributes $\mathbf{A}=[a_i]$, where $a_i$ is the real-valued attribute (or vector of attributes) of $v_i$.  (These may be, for example, the list of refined vertices $\mathbf{V}_r$ and their associated colors $\mathbf{C}_r=\mathbf{C}$ as discussed above.)  In the process of {\em voxelization}, the points are partitioned into voxels, and the attributes associated with the points in a voxel are averaged.  The points within each voxel are quantized to the voxel center.  Each occupied voxel is then represented by the voxel center and the average of the attributes of the points in the voxel.  Moreover, the occupied voxels are put into Z-scan order, also known as Morton order \cite{Morton66}.  The first step in voxelization is to quantize the vertices and to produce their Morton codes.  The Morton code $m$ for a point $(x,y,z)$ is obtained simply by interleaving (or ``swizzling'') the bits of $x$, $y$, and $z$, with $x$ being higher order than $y$, and $y$ being higher order than $z$.  For example, if  $x=x_4x_2x_1$, $y=y_4y_2y_1$, and $z=z_4z_2z_1$ (written in binary), then the Morton code for the point would be $m=x_4y_4z_4x_2y_2z_2x_1y_1z_1$.  The Morton codes are sorted, duplicates are removed, and all attributes whose vertices have a particular Morton code are averaged.


 
The procedure is detailed in Algorithm~\ref{alg:voxelize}.  $\mathbf{V}_{int}$ is the list of vertices with their coordinates, previously in $[0,1)$, now mapped to integers in $\{0,\ldots,2^J-1\}$.  $\mathbf{M}$ is the corresponding list of Morton codes.  $\mathbf{M}_v$ is the list of Morton codes, sorted with duplicates removed, using the Matlab function {\em unique}.  $\mathbf{I}$ and $\mathbf{I}_v$ are vectors of indices such that $\mathbf{M}_v=\mathbf{M}(\mathbf{I})$ and $\mathbf{M}=\mathbf{M}_v(\mathbf{I}_v)$, in Matlab notation. (That is, the $i_v$th element of $\mathbf{M}_v$ is the $\mathbf{I}(i_v)$th element of $\mathbf{M}$ and the $i$th element of $\mathbf{M}$ is the $\mathbf{I}_v(i)$th element of $\mathbf{M}_v$.) $\mathbf{A}_v=[\bar a_j]$ is the list of attribute averages
\begin{equation}
\bar{a}_j =\frac{1}{N_j} \sum_{i:\mathbf{M}(i)=\mathbf{M}_v(j)} a_i,
\end{equation}
where $N_j$ is the number of elements in the sum.  $\mathbf{V}_v$ is the list of voxel centers.  The algorithm has complexity $\mathcal{O}\left(N\log N\right)$, where $N$ is the number of input vertices.

\begin{algorithm}
\caption{Voxelization ({\em voxelize})}\label{alg:voxelize}
\begin{algorithmic}[1]
\Require $\mathbf{V}$, $\mathbf{A}$, $J$
\State $\mathbf{V}_{int}=floor(\mathbf{V}*2^J)$ // map coords to $\{0,\ldots,2^J-1\}$
\State $\mathbf{M}=morton(\mathbf{V}_{int})$ // generate list of morton codes
\State $[\mathbf{M}_v,\mathbf{I},\mathbf{I}_v]=unique(\mathbf{M})$ // find unique codes, and sort
\State $\mathbf{A}_v=[\bar{a}_j]$, where $\bar{a}_j=mean(\mathbf{A}(\mathbf{M}=\mathbf{M}_v(j))$ is the average of all attributes whose Morton code is the $j$th Morton code in the list $\mathbf{M}_v$
\State $\mathbf{V}_v=(\mathbf{V}_{int}(\mathbf{I},:)+0.5)*2^{-J}$ // compute voxel centers
\Ensure $\mathbf{V}_v$ (or equivalently $\mathbf{M}_v$), $\mathbf{A}_v$, $\mathbf{I}_v$.
\end{algorithmic}
\end{algorithm}


\subsection{Octree encoding}

\begin{figure}[t]
\centering
\includegraphics[scale=0.47]{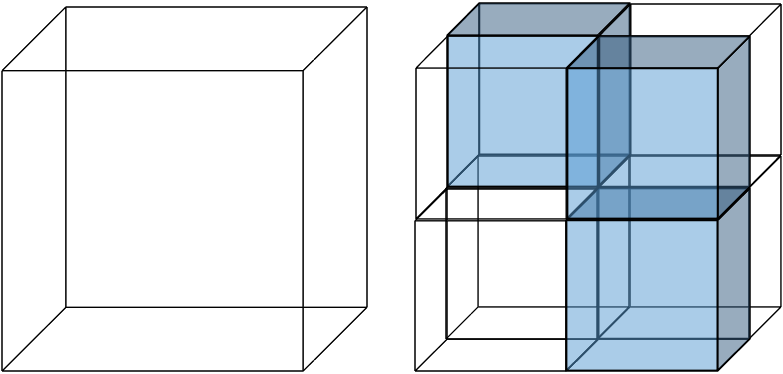}
\caption{Cube subdivision. Blue cubes represent occupied regions of space.}
\label{fig_voxel_division}
\end{figure}

Any set of voxels in the unit cube, each of size $2^{-J} \times 2^{-J} \times 2^{-J}$, designated {\em occupied} voxels, can be represented with an octree of depth $J$ \cite{JackinsT80,meagher_octtree}.  An octree is a recursive subdivision of a cube into smaller cubes, as illustrated in Figure~\ref{fig_voxel_division}.  Cubes are subdivided only as long as they are occupied (i.e., contain any occupied voxels).
This recursive subdivision can be represented by an octree with depth $J$, where the root corresponds to the unit cube. The leaves of the tree correspond to the set of occupied voxels.

There is a close connection between octrees and Morton codes.  In fact, the Morton code of a voxel, which has length $3J$ bits broken into $J$ binary triples, encodes the path in the octree from the root to the leaf containing the voxel.  Moreover, the sorted list of Morton codes results from a depth-first traversal of the tree.

Each internal node of the tree can be represented by one byte, to indicate which of its eight children are occupied.  If these bytes are serialized in a depth-first traversal of the tree, the serialization (which has a length in bytes equal to the number of internal nodes of the tree) can be used as a description of the octree, from which the octree can be reconstructed.  Hence the description can also be used to encode the ordered list of Morton codes of the leaves.  This description can be further compressed using a context adaptive arithmetic encoder.  However, for simplicity in our experiments, we use {\em gzip} instead of an arithmetic encoder.

In this way, we encode any set of occupied voxels in a canonical (Morton) order.

\subsection{Transform coding}
\label{subsec:transform_coding}

In this section we describe the region adaptive hierarchical transform (RAHT) \cite{queiroz_raht} and its efficient implementation.  RAHT can be described as a sequence of orthonormal transforms applied to attribute data living on the leaves of an octree. For simplicity we assume the attributes are scalars.  This transform processes voxelized attributes in a bottom up fashion, starting at the leaves of the octree.  The inverse transform reverses this order.

\begin{figure}[t]
\centering
\includegraphics[scale=0.47]{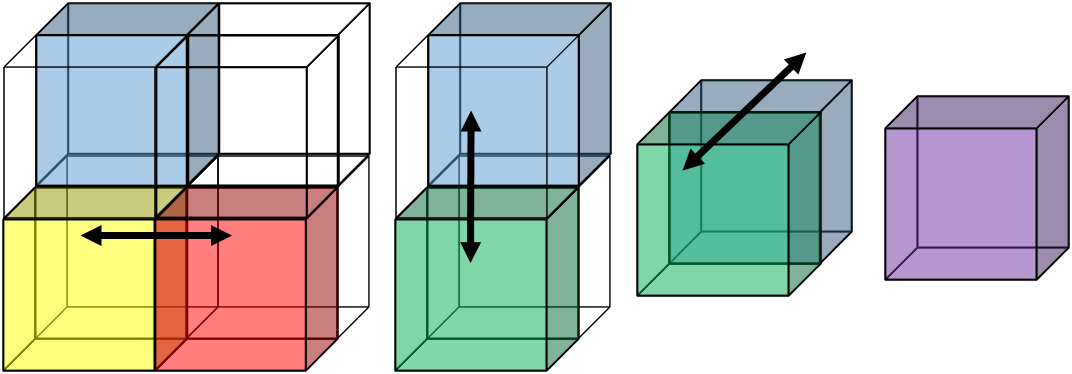}
\caption{One level of  RAHT applied to a cube of eight voxels, three of which are occupied.}
\label{fig_RAHT_application}
\end{figure}

Consider eight adjacent voxels, three of which are occupied, having the same parent in the octree, as shown in Figure~\ref{fig_RAHT_application}.  The colored voxels are occupied (have an attribute) and the transparent ones are empty.  Each occupied voxel is assigned a unit weight.  For the forward transform, transformed attribute values and weights will be propagated up the tree.

One level of the forward transform proceeds as follows.  Pick a direction ($x,y,z$), then check whether there are two occupied cubes that can be processed along that direction. In the leftmost part of Figure~\ref{fig_RAHT_application} there are only three occupied cubes, \emph{red}, \emph{yellow}, and \emph{blue}, having weights $w_{r}$, $w_{y}$, and $w_{b}$, respectively.  To process in the direction of the $x$ axis, since the \emph{blue} cube  does not have a neighbor along the horizontal direction, we copy its attribute value $a_b$ to the second stage and keep its weight $w_{b}$. The attribute values $a_y$ and $a_r$ of the \emph{yellow} and \emph{red} cubes can be processed together using the orthonormal transformation
\begin{align}\label{eq_raht}
\begin{bmatrix}
a_g^0 \\a_g^1
\end{bmatrix}=
\frac{1}{\sqrt{w_y+w_r}}\begin{bmatrix}
\sqrt{w_y} & \sqrt{w_r} \\
-\sqrt{w_r} & \sqrt{w_y}
\end{bmatrix}
\begin{bmatrix}
a_y \\a_r 
\end{bmatrix},
\end{align}
where the transformed coefficients $a_g^0$ and $a_g^1$ respectively represent low pass and high pass coefficients appropriately weighted.  Both transform coefficients now represent information from a region with weight $w_g=w_y+w_r$ (\emph{green} cube). The high pass coefficient is stored for entropy coding along with its weight, while the low pass coefficient is further processed and put in the \emph{green} cube. For processing along the $y$ axis, the \emph{green} and \emph{blue} cubes do not have neighbors, so their values are copied to the next level. Then we process in the $z$ direction using the same transformation in (\ref{eq_raht}) with weights $w_g$ and $w_b$.

This process is repeated for each cube of eight subcubes at each level of the octree. After $J$ levels, there remains one low pass coefficient that corresponds to the DC component; the remainder are high pass coefficients.  Since after each processing of a pair of coefficients, the weights are added and used during the next transformation, the weights can be interpreted as being inversely proportional to frequency.  The DC coefficient is the one that has the largest weight, as it is processed more times and represents information from the entire cube, while the high pass coefficients, which are produced earlier, have smaller weights because they contain information from a smaller region. The weights depend only on the octree (not the coefficients themselves), and thus can provide a frequency ordering for the coefficients. We sort the transformed coefficients by decreasing magnitude of weight.

\begin{figure}
\includegraphics[width=0.48\textwidth]{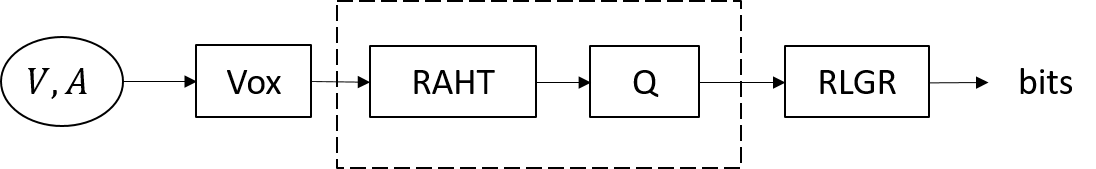}
\caption{Transform coding system for voxelized point clouds.}
\label{fig:vox_RAHT_Q_RLGR}
\end{figure}

Finally, the sorted coefficients are quantized using uniform scalar quantization, and are entropy coded using adaptive Run Length Golomb-Rice coding \cite{malvar_rlgr}.  The pipeline is illustrated in Figure~\ref{fig:vox_RAHT_Q_RLGR}.

Efficient implementations of RAHT and its inverse are detailed in Algorithms~\ref{alg:RAHT} and~\ref{alg:IRAHT}, respectively.  Algorithm~\ref{alg:prologue} is a prologue to each.  Algorithm~\ref{alg:quantize} is our uniform scalar quantization.

\begin{algorithm}
\caption{Prologue to Region Adaptive Hierarchical Transform (RAHT) and its Inverse (IRAHT) ({\em prologue})}
\label{alg:prologue}
\begin{algorithmic}[1]
\Require $\mathbf{V}$, $J$
\State $\mathbf{M}_1=morton(\mathbf{V})$ // morton codes
\State $N=length(\mathbf{M}_1)$ // number of points
\For{$\ell=1$ to $3J$} // define $(\mathbf{I}_\ell,\mathbf{M}_\ell,\mathbf{W}_\ell,\mathbf{F}_\ell), \forall\ell$
	\If{$\ell=1$} // initialize indices of coeffs at layer 1
    	\State $\mathbf{I}_1 = (1:N)^T$ // vector of indices from 1 to $N$
    \Else \ // define indices of coeffs at layer $\ell$
        \State $\mathbf{I}_\ell=\mathbf{I}_{\ell-1}(\lnot[0;\mathbf{F}_{\ell-1}])$ // left sibs and singletons
    \EndIf
    \State $\mathbf{M}_\ell=\mathbf{M}_1(\mathbf{I}_\ell)$ // morton codes at layer $\ell$
	\State $\mathbf{W}_\ell=[\mathbf{I}_\ell(2:end);N+1]-\mathbf{I}_\ell$ // weights
    \State $\mathbf{D}=\mathbf{M}_\ell(1:end-1) \oplus \mathbf{M}_\ell(2:end)$ // path diffs
    \State $\mathbf{F}_\ell=(\mathbf{D} \wedge (2^{3J}-2^\ell))=0$ // left sibling flags
\EndFor
\Ensure $\{(\mathbf{I}_\ell,\mathbf{W}_\ell,\mathbf{F}_\ell):\ell=1,\ldots,3J\}$, and $N$
\end{algorithmic}
\end{algorithm}

\begin{algorithm}
\caption{Region Adaptive Hierarchical Transform ({\em RAHT})}
\label{alg:RAHT}
\begin{algorithmic}[1]
\Require $\mathbf{V}$, $\mathbf{A}$, $J$
\State $[\{(\mathbf{I}_\ell,\mathbf{W}_\ell,\mathbf{F}_\ell)\},N]=prologue(\mathbf{V},J)$
\State $\mathbf{TA}=\mathbf{A}$ // perform transform in place
\State $\mathbf{W}=\mathbf{1}$ // initialize to $N$-vector of unit weights
\For{$\ell=1$ to $3J-1$}
	\State $\mathbf{i}_0=\mathbf{I}_\ell([\mathbf{F}_\ell;0]==1)$ // left sibling indices
	\State $\mathbf{i}_1=\mathbf{I}_\ell([0;\mathbf{F}_\ell]==1)$ // right sibling indices
    \State $\mathbf{w}_0=\mathbf{W}_\ell([\mathbf{F}_\ell;0]==1)$ // left sibling weights
	\State $\mathbf{w}_1=\mathbf{W}_\ell([0;\mathbf{F}_\ell]==1)$ // right sibling weights
    \State $\mathbf{x}_0=\mathbf{TA}(\mathbf{i}_0,:)$ // left sibling coefficients
    \State $\mathbf{x}_1=\mathbf{TA}(\mathbf{i}_1,:)$ // right sibling coefficients
    \State $\mathbf{a}=repmat(sqrt(\mathbf{w}_0 ./ (\mathbf{w}_0+\mathbf{w}_1)),1,size(\mathbf{TA},2))$
    \State $\mathbf{b}=repmat(sqrt(\mathbf{w}_1 ./ (\mathbf{w}_0+\mathbf{w}_1)),1,size(\mathbf{TA},2))$
    \State $\mathbf{TA}(\mathbf{i}_0,:)=\mathbf{a}\ .* \mathbf{x}_0 + \mathbf{b}\ .* \mathbf{x}_1$
    \State $\mathbf{TA}(\mathbf{i}_1,:)=-\mathbf{b}\ .* \mathbf{x}_0 + \mathbf{a}\ .* \mathbf{x}_1$
    \State $\mathbf{W}(\mathbf{i}_0)=\mathbf{W}(\mathbf{i}_0)+\mathbf{W}(\mathbf{i}_1)$
    \State $\mathbf{W}(\mathbf{i}_1)=\mathbf{W}(\mathbf{i}_0)$
\EndFor
\Ensure $\mathbf{TA}$, $\mathbf{W}$
\end{algorithmic}
\end{algorithm}

\begin{algorithm}
\caption{Inverse Region Adaptive Hierarchical Transform ({\em IRAHT})}
\label{alg:IRAHT}
\begin{algorithmic}[1]
\Require $\mathbf{V}$, $\mathbf{TA}$, $J$
\State $[\{(\mathbf{I}_\ell,\mathbf{W}_\ell,\mathbf{F}_\ell)\},N]=prologue(\mathbf{V},J)$
\State $\mathbf{A}=\mathbf{TA}$ // perform inverse transform in place
\For{$\ell=3J-1$ down to $1$}
	\State $\mathbf{i}_0=\mathbf{I}_\ell([\mathbf{F}_\ell;0]==1)$ // left sibling indices
	\State $\mathbf{i}_1=\mathbf{I}_\ell([0;\mathbf{F}_\ell]==1)$ // right sibling indices
    \State $\mathbf{w}_0=\mathbf{W}_\ell([\mathbf{F}_\ell;0]==1)$ // left sibling weights
	\State $\mathbf{w}_1=\mathbf{W}_\ell([0;\mathbf{F}_\ell]==1)$ // right sibling weights
    \State $\mathbf{x}_0=\mathbf{TA}(\mathbf{i}_0,:)$ // left sibling coefficients
    \State $\mathbf{x}_1=\mathbf{TA}(\mathbf{i}_1,:)$ // right sibling coefficients
    \State $\mathbf{a}=repmat(sqrt(\mathbf{w}_0 ./ (\mathbf{w}_0+\mathbf{w}_1)),1,size(\mathbf{TA},2))$
    \State $\mathbf{b}=repmat(sqrt(\mathbf{w}_1 ./ (\mathbf{w}_0+\mathbf{w}_1)),1,size(\mathbf{TA},2))$
    \State $\mathbf{TA}(\mathbf{i}_0,:)=\mathbf{a}\ .* \mathbf{x}_0 - \mathbf{b}\ .* \mathbf{x}_1$
    \State $\mathbf{TA}(\mathbf{i}_1,:)=\mathbf{b}\ .* \mathbf{x}_0 + \mathbf{a}\ .* \mathbf{x}_1$
\EndFor
\Ensure $\mathbf{A}$
\end{algorithmic}
\end{algorithm}

\begin{algorithm}
\caption{Uniform scalar quantization ({\em quantize})}
\label{alg:quantize}
\begin{algorithmic}[1]
\Require $\mathbf{A}$, $step$, $midriseORmidstep$
\If{$midriseORmidstep = midstep$}
\State $\hat{\mathbf{A}}=round(\mathbf{A}/step) * step$
\Else \ // $midriseORmidstep = midrise$
\State $\hat{\mathbf{A}}=[round(\mathbf{A}/step-0.5) + 0.5] * step$
\EndIf
\Ensure $\hat{\mathbf{A}}$
\end{algorithmic}
\end{algorithm}

\section{Encoding and Decoding}
\label{sec:system}
\begin{figure*}[t]
\centering
\includegraphics[scale=0.48]{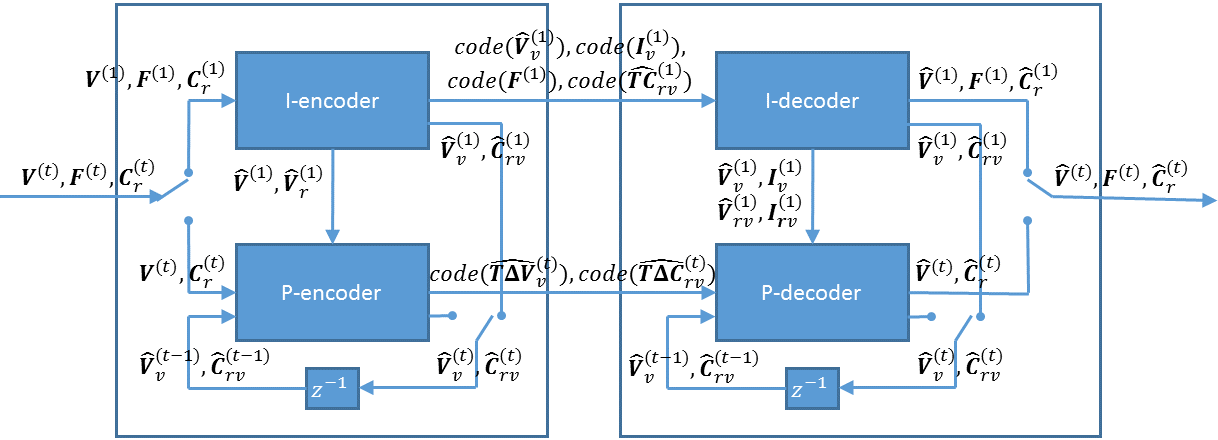}
\caption{Encoder (left) and decoder (right).  The switches are in the $t=1$ position, and flip for $t>1$.}
\label{fig:systemoverview}
\end{figure*}

In this section we describe in detail encoding and decoding of dynamic triangle clouds.
First we describe encoding and decoding of reference frames.
Following that, we describe encoding and decoding of predicted frames.
For both reference and predicted frames, we describe first how geometry is encoded and decoded, and then how color is encoded and decoded.
The overall system is shown in Figure~\ref{fig:systemoverview}.

\subsection{Encoding and decoding of reference frames}

For reference frames, encoding is summarized in Algorithm~\ref{alg:Iencoder}, while decoding is summarized in Algorithm~\ref{alg:Idecoder}.

\begin{algorithm}
\caption{Encode reference frame (I-encoder)}
\label{alg:Iencoder}
\begin{algorithmic}[1]
\Require $J$, $U$, $\Delta_{color,intra}$ (from system parameters)
\Require $\mathbf{V}^{(1)}$, $\mathbf{F}^{(1)}$, $\mathbf{C}_r^{(1)}$ (from system input)
\State // Geometry
\State $\hat{\mathbf{V}}^{(1)}=quantize(\mathbf{V}^{(1)},2^{-J},midrise)$
\State $[\hat{\mathbf{V}}_v^{(1)},\mathbf{V}_v^{(1)},\mathbf{I}_v^{(1)}]=voxelize(\hat{\mathbf{V}}^{(1)},\mathbf{V}^{(1)},J)$ s.t. $\hat{\mathbf{V}}^{(1)}$ $=$ $\hat{\mathbf{V}}_v^{(1)}(\mathbf{I}_v^{(1)})$
\State // Color
\State $\hat{\mathbf{V}}_r^{(1)}=refine(\hat{\mathbf{V}}^{(1)},\mathbf{F}^{(1)},U)$
\State $[\hat{\mathbf{V}}_{rv}^{(1)},\mathbf{C}_{rv}^{(1)},\mathbf{I}_{rv}^{(1)}]=voxelize(\hat{\mathbf{V}}_r^{(1)},\mathbf{C}_r^{(1)},J)$ s.t. $\hat{\mathbf{V}}_r^{(1)}=\hat{\mathbf{V}}_{rv}^{(1)}(\mathbf{I}_{rv}^{(1)})$
\State $[\mathbf{TC}_{rv}^{(1)},\mathbf{W}_{rv}^{(1)}]=RAHT(\hat{\mathbf{V}}_{rv}^{(1)},\mathbf{C}_{rv}^{(1)},J)$
\State $\widehat{\mathbf{TC}}_{rv}^{(1)}=quantize(\mathbf{TC}_{rv}^{(1)},\Delta_{color,intra},midstep)$
\State $\hat{\mathbf{C}}_{rv}^{(1)}=IRAHT(\hat{\mathbf{V}}_{rv}^{(1)},\widehat{\mathbf{TC}}_{rv}^{(1)},J)$
\Ensure $code(\hat{\mathbf{V}}_v^{(1)})$, $code(\mathbf{I}_v^{(1)})$, $code(\mathbf{F}^{(1)})$, $code(\widehat{\mathbf{TC}}_{rv}^{(1)})$ (to reference frame decoder)
\Ensure $\hat{\mathbf{V}}^{(1)}$, $\hat{\mathbf{V}}_r^{(1)}$ (to predicted frame encoder)
\Ensure $\hat{\mathbf{V}}_v^{(1)}$, $\hat{\mathbf{C}}_{rv}^{(1)}$ (to reference frame buffer)
\end{algorithmic}
\end{algorithm}

\begin{algorithm}
\caption{Decode reference frame (I-decoder)}
\label{alg:Idecoder}
\begin{algorithmic}[1]
\Require $J$, $U$, $\Delta_{color,intra}$ (from system parameters)
\Require $code(\hat{\mathbf{V}}_v^{(1)})$, $code(\mathbf{I}_v^{(1)})$, $code(\mathbf{F}^{(1)})$, $code(\widehat{\mathbf{TC}}_{rv}^{(1)})$ (from reference frame encoder)
\State // Geometry
\State $\hat{\mathbf{V}}^{(1)}=\hat{\mathbf{V}}_v^{(1)}(\mathbf{I}_v^{(1)})$
\State // Color
\State $\hat{\mathbf{V}}_r^{(1)}=refine(\hat{\mathbf{V}}^{(1)},\mathbf{F}^{(1)},U)$
\State $[\hat{\mathbf{V}}_{rv}^{(1)},\mathbf{I}_{rv}^{(1)}]=voxelize(\hat{\mathbf{V}}_r^{(1)},J)$ s.t. $\hat{\mathbf{V}}_r^{(1)}=\hat{\mathbf{V}}_{rv}^{(1)}(\mathbf{I}_{rv}^{(1)})$
\State $\mathbf{W}_{rv}^{(1)}=RAHT(\hat{\mathbf{V}}_{rv}^{(1)},J)$
\State $\hat{\mathbf{C}}_{rv}^{(1)}=IRAHT(\hat{\mathbf{V}}_{rv}^{(1)},\widehat{\mathbf{TC}}_{rv}^{(1)},J)$
\State $\hat{\mathbf{C}}_r^{(1)}=\hat{\mathbf{C}}_{rv}^{(1)}(\mathbf{I}_{rv}^{(1)})$
\Ensure $\hat{\mathbf{V}}^{(1)}$, $\mathbf{F}^{(1)}$, $\hat{\mathbf{C}}_r^{(1)}$ (to renderer)
\Ensure $\hat{\mathbf{V}}_v^{(1)}$, $\mathbf{I}_v^{(1)}$, $\hat{\mathbf{V}}_{rv}^{(1)}$, $\mathbf{I}_{rv}^{(1)}$ (to predicted frame decoder)
\Ensure $\hat{\mathbf{V}}_v^{(1)}$, $\hat{\mathbf{C}}_{rv}^{(1)}$ (to reference frame buffer)
\end{algorithmic}
\end{algorithm}

\subsubsection{Geometry encoding and decoding}
\label{subsec:reference_geometry}

We assume that the vertices in $\mathbf{V}^{(1)}$ are in Morton order.  If not, we put them into Morton order and permute the indices in $\mathbf{F}^{(1)}$ accordingly.  The lists $\mathbf{V}^{(1)}$ and $\mathbf{F}^{(1)}$ are the geometry-related quantities in the reference frame transmitted from the encoder to the decoder.  $\mathbf{V}^{(1)}$ will be reconstructed at the decoder with some loss as $\hat{\mathbf{V}}^{(1)}$, and $\mathbf{F}^{(1)}$ will be reconstructed losslessly.  We now describe the process.

At the encoder, the vertices in $\mathbf{V}^{(1)}$ are first quantized to the voxel grid, producing a list of quantized vertices $\hat{\mathbf{V}}^{(1)}$, the same length as $\mathbf{V}^{(1)}$.  There may be duplicates in $\hat{\mathbf{V}}^{(1)}$, because some vertices may have collapsed to the same grid point.  $\hat{\mathbf{V}}^{(1)}$ is then voxelized (without attributes), the effect of which is simply to remove the duplicates, producing a possibly slightly shorter list $\hat{\mathbf{V}}_v^{(1)}$ along with a list of indices $\mathbf{I}_v^{(1)}$ such that (in Matlab notation) $\hat{\mathbf{V}}^{(1)}=\hat{\mathbf{V}}_v^{(1)}(\mathbf{I}_v^{(1)})$.  Since $\hat{\mathbf{V}}_v^{(1)}$ has no duplicates, it represents a {\em set} of voxels.  This set can be described by an octree.  The byte sequence representing the octree can be compressed with any entropy encoder; we use {\em gzip}.  The list of indices $\mathbf{I}_v^{(1)}$, which has the same length as $\hat{\mathbf{V}}^{(1)}$, indicates, essentially, how to restore the duplicates, which are missing from $\hat{\mathbf{V}}_v^{(1)}$.  In fact, the indices in $\mathbf{I}_v^{(1)}$ increase in unit steps for all vertices in $\hat{\mathbf{V}}^{(1)}$ except the duplicates, for which there is no increase.  The list of indices is thus a sequence of runs of unit increases alternating with runs of zero increases.  This binary sequence of increases can be encoded with any entropy encoder; we use {\em gzip} on the run lengths.  Finally the list of faces $\mathbf{F}^{(1)}$ can be encoded with any entropy encoder; we again use {\em gzip}, though algorithms such as \cite{Rossignac_1999,MamouZP09} might also be used.

The decoder entropy decodes $\hat{\mathbf{V}}_v^{(1)}$, $\mathbf{I}_v^{(1)}$, and $\mathbf{F}^{(1)}$, and then recovers $\hat{\mathbf{V}}^{(1)}=\hat{\mathbf{V}}_v^{(1)}(\mathbf{I}_v^{(1)})$, which is the quantized version of $\mathbf{V}^{(1)}$, to obtain both $\hat{\mathbf{V}}^{(1)}$ and $\mathbf{F}^{(1)}$.

\subsubsection{Color encoding and decoding}

Let $\mathbf{V}_r^{(1)}=refine(\mathbf{V}^{(1)},\mathbf{F}^{(1)},U)$ be the list of ``refined vertices'' obtained by upsampling, by factor $U$, the faces $\mathbf{F}^{(1)}$ whose vertices are $\mathbf{V}^{(1)}$.  We assume that the colors in the list $\mathbf{C}_r^{(1)}=\mathbf{C}^{(1)}$ correspond to the refined vertices in $\mathbf{V}_r^{(1)}$.  In particular, the lists have the same length.  Here, we subscript the list of colors by an `r' to indicate that it corresponds to the list of refined vertices.

When the vertices $\mathbf{V}^{(1)}$ are quantized to $\hat{\mathbf{V}}^{(1)}$, the refined vertices change to $\hat{\mathbf{V}}_r^{(1)}=refine(\hat{\mathbf{V}}^{(1)},\mathbf{F}^{(1)},U)$.  The list of colors $\mathbf{C}_r^{(1)}$ can also be considered as indicating the colors on $\hat{\mathbf{V}}_r^{(1)}$.   The list $\mathbf{C}_r^{(1)}$ is the color-related quantity in the reference frame transmitted from the encoder to the decoder.  The decoder will reconstruct $\mathbf{C}_r^{(1)}$ with some loss $\hat{\mathbf{C}}_r^{(1)}$.  We now describe the process.

At the encoder, the refined vertices $\hat{\mathbf{V}}_r^{(1)}$ are obtained as described above.  Then the vertices $\hat{\mathbf{V}}_r^{(1)}$ and their associated color attributes $\mathbf{C}_r^{(1)}$ are voxelized to obtain a list of voxels $\hat{\mathbf{V}}_{rv}^{(1)}$, the list of voxel colors  $\mathbf{C}_{rv}^{(1)}$, and the list of indices $\mathbf{I}_{rv}^{(1)}$ such that (in Matlab notation) $\hat{\mathbf{V}}_r^{(1)}=\hat{\mathbf{V}}_{rv}^{(1)}(\mathbf{I}_{rv}^{(1)})$.  The list of indices $\mathbf{I}_{rv}^{(1)}$ has the same length as $\hat{\mathbf{V}}_r^{(1)}$, and contains for each vertex in $\hat{\mathbf{V}}_r^{(1)}$ the index of its corresponding vertex in $\hat{\mathbf{V}}_{rv}^{(1)}$.  Particularly if the upsampling factor $U$ is large, there may be many refined vertices falling into each voxel.  Hence the list $\hat{\mathbf{V}}_{rv}^{(1)}$ may be significantly shorter than the list $\hat{\mathbf{V}}_r^{(1)}$ (and the list $\mathbf{I}_{rv}^{(1)}$).  However, unlike the geometry case, in this case the list $\mathbf{I}_{rv}^{(1)}$ need not be transmitted.

The list of voxel colors $\hat{\mathbf{C}}_{rv}^{(1)}$, each with unit weight, is transformed by RAHT to an equal-length list of transformed colors $\mathbf{TC}_{rv}^{(1)}$ and associated weights $\mathbf{W}_{rv}^{(1)}$.  The transformed colors then quantized with stepsize $\Delta_{color,intra}$ to obtain $\widehat{\mathbf{TC}}_{rv}^{(1)}$.  The quantized RAHT coefficients are entropy coded
as described in Section~\ref{subsec:transform_coding}
using the associated weights, and are transmitted.  Finally, $\widehat{\mathbf{TC}}_{rv}^{(1)}$ is inverse transformed by RAHT to obtain $\hat{\mathbf{C}}_{rv}^{(1)}$.  These represent the quantized voxel colors, and will be used as a reference for subsequent predicted frames.

At the decoder, similarly, the refined vertices $\hat{\mathbf{V}}_r^{(1)}$ are obtained by upsampling, by factor $U$, the faces $\mathbf{F}^{(1)}$ whose vertices are $\hat{\mathbf{V}}^{(1)}$ (both of which have been decoded already in the geometry step).  $\hat{\mathbf{V}}_r^{(1)}$ is then voxelized (without attributes) to produce the list of voxels $\hat{\mathbf{V}}_{rv}^{(1)}$ and list of indices $\mathbf{I}_{rv}^{(1)}$ such that $\hat{\mathbf{V}}_r^{(1)}=\hat{\mathbf{V}}_{rv}^{(1)}(\mathbf{I}_{rv}^{(1)})$. The weights $\mathbf{W}_{rv}^{(1)}$ are recovered by using RAHT to transform a null signal on the vertices $\hat{\mathbf{V}}_r^{(1)}$, each with unit weight.  Then $\widehat{\mathbf{TC}}_{rv}^{(1)}$ is entropy decoded using the recovered weights and inverse transformed by RAHT to obtain the quantized voxel colors $\hat{\mathbf{C}}_{rv}^{(1)}$.  Finally, the quantized refined vertex colors can be obtained as $\hat{\mathbf{C}}_r^{(1)}=\hat{\mathbf{C}}_{rv}^{(1)}(\mathbf{I}_{rv}^{(1)})$.

\subsection{Encoding and decoding of predicted frames}


We assume that all $N$ frames in a GOP are aligned.  That is, the lists of faces, $\mathbf{F}^{(1)},\ldots,\mathbf{F}^{(N)}$, are all identical.  Moreover, the lists of vertices, $\mathbf{V}^{(1)},\ldots,\mathbf{V}^{(N)}$, all correspond in the sense that the $i$th vertex in list $\mathbf{V}^{(1)}$ (say, $v^{(1)}(i)=v_i^{(1)}$) corresponds to the $i$th vertex in list $\mathbf{V}^{(t)}$ (say, $v^{(t)}(i)=v_i^{(t)}$), for all $t=1,\ldots,N$.  $(v^{(1)}(i),\ldots,v^{(N)}(i))$ is the trajectory of vertex $i$ over the GOF, $i=1,\ldots,N_p$, where $N_p$ is the number of vertices.

Similarly, when the faces are upsampled by factor $U$ to create new lists of refined vertices, $\mathbf{V}_r^{(1)},\ldots,\mathbf{V}_r^{(N)}$ --- and their colors, $\mathbf{C}_r^{(1)},\ldots,\mathbf{C}_r^{(N)}$ --- the $i_r$th elements of these lists also correspond to each other across the GOF, $i_r=1,\ldots,N_c$, where $N_c$ is the number of refined vertices, or the number of colors.

The trajectory $\{(v^{(1)}(i),\ldots,v^{(N)}(i)):i=1,\ldots,N_p\}$ can be considered an attribute of vertex $v^{(1)}(i)$, and likewise the trajectories $\{(v_r^{(1)}(i_r),\ldots,v_r^{(N)}(i_r)):i_r=1,\ldots,N_c\}$ and $\{(c_r^{(1)}(i_r),\ldots,c_r^{(N)}(i_r)):i_r=1,\ldots,N_c\}$ can be considered attributes of refined vertex $v_r^{(1)}(i_r)$.  Thus the trajectories can be partitioned according to how the vertex $v^{(1)}(i)$ and the refined vertex $v_r^{(1)}(i_r)$ are voxelized.  As for any attribute, the average of the trajectories in each cell of the partition is used to represent all trajectories in the cell.  Our scheme codes these representative trajectories.  This could be a problem if trajectories diverge from the same, or nearly the same, point, for example, when clapping hands separate.  However, this situation is usually avoided by restarting the GOF by inserting a key frame, or reference frame, whenever the topology changes, and by using a sufficiently fine voxel grid.

In this section we show how to encode and decode the predicted frames, i.e., frames $t=2,\ldots,N$, in each GOF.  The frames are processed one at a time, with no look-ahead, to minimize latency. The encoding is detailed in Algorithm~\ref{alg:Pencoder}, while decoding is detailed in Algorithm~\ref{alg:Pdecoder}.

\begin{algorithm}
\caption{Encode predicted frame (P-encoder)}
\label{alg:Pencoder}
\begin{algorithmic}[1]
\Require $J$, $\Delta_{motion}$, $\Delta_{color,inter}$ (from system parameters)
\Require $\mathbf{V}^{(t)}$, $\mathbf{C}_r^{(t)}$ (from system input)
\Require $\hat{\mathbf{V}}^{(1)},\hat{\mathbf{V}}_r^{(1)}$ (from reference frame encoder)
\Require $\hat{\mathbf{V}}_v^{(t-1)}$, $\hat{\mathbf{C}}_{rv}^{(t-1)}$ (from previous frame buffer)
\State // Geometry
\State $[\hat{\mathbf{V}}_v^{(1)},\mathbf{V}_v^{(t)},\mathbf{I}_v^{(1)}]=voxelize(\hat{\mathbf{V}}^{(1)},\mathbf{V}^{(t)},J)$ s.t. $\hat{\mathbf{V}}^{(1)}=\hat{\mathbf{V}}_v^{(1)}(\mathbf{I}_v^{(1)})$
\State $\Delta\mathbf{V}_v^{(t)}=\mathbf{V}_v^{(t)}-\hat{\mathbf{V}}_v^{(t-1)}$
\State $[\mathbf{T}\Delta\mathbf{V}_v^{(t)},\mathbf{W}_v^{(1)}]=RAHT(\hat{\mathbf{V}}_v^{(1)},\Delta\mathbf{V}_v^{(t)},J)$
\State $\widehat{\mathbf{T}\Delta\mathbf{V}}_v^{(t)}=quantize(\mathbf{T}\Delta\mathbf{V}_v^{(t)},\Delta_{motion},midstep)$
\State $\widehat{\Delta\mathbf{V}}_v^{(t)}=IRAHT(\hat{\mathbf{V}}_v^{(1)},\widehat{\mathbf{T}\Delta\mathbf{V}}_v^{(t)},J)$
\State $\hat{\mathbf{V}}_v^{(t)}=\hat{\mathbf{V}}_v^{(t-1)}+\widehat{\Delta\mathbf{V}}_v^{(t)}$
\State // Color
\State $[\hat{\mathbf{V}}_{rv}^{(1)},\mathbf{C}_{rv}^{(t)},\mathbf{I}_{rv}^{(1)}]=voxelize(\hat{\mathbf{V}}_r^{(1)},\mathbf{C}_r^{(t)},J)$ s.t. $\hat{\mathbf{V}}_r^{(1)}=\hat{\mathbf{V}}_{rv}^{(1)}(\mathbf{I}_{rv}^{(1)})$
\State $\Delta\mathbf{C}_{rv}^{(t)}=\mathbf{C}_{rv}^{(t)}-\hat{\mathbf{C}}_{rv}^{(t-1)}$
\State $[\mathbf{T}\Delta\mathbf{C}_{rv}^{(t)},\mathbf{W}_{rv}^{(1)}]=RAHT(\hat{\mathbf{V}}_{rv}^{(1)},\Delta\mathbf{C}_{rv}^{(t)},J)$
\State $\widehat{\mathbf{T}\Delta\mathbf{C}}_{rv}^{(t)}=quantize(\mathbf{T}\Delta\mathbf{C}_{rv}^{(t)},\Delta_{color,inter},midstep)$
\State $\widehat{\Delta\mathbf{C}}_{rv}^{(t)}=IRAHT(\hat{\mathbf{V}}_{rv}^{(1)},\widehat{\mathbf{T}\Delta\mathbf{C}}_{rv}^{(t)},J)$
\State $\hat{\mathbf{C}}_{rv}^{(t)}=\hat{\mathbf{C}}_{rv}^{(t-1)}+\widehat{\Delta\mathbf{C}}_{rv}^{(t)}$
\Ensure $code(\widehat{\mathbf{T}\Delta\mathbf{V}}_v^{(t)})$, $code(\widehat{\mathbf{T}\Delta\mathbf{C}}_{rv}^{(t)})$ (to predicted frame decoder)
\Ensure $\hat{\mathbf{V}}_v^{(t)}$, $\hat{\mathbf{C}}_{rv}^{(t)}$ (to previous frame buffer)
\end{algorithmic}
\end{algorithm}

\begin{algorithm}
\caption{Decode predicted frame (P-decoder)}
\label{alg:Pdecoder}
\begin{algorithmic}[1]
\Require $J$, $U$, $\Delta_{motion}$, $\Delta_{color,inter}$ (from system parameters)
\Require $code(\widehat{\mathbf{T}\Delta\mathbf{V}}_v^{(t)})$, $code(\widehat{\mathbf{T}\Delta\mathbf{C}}_{rv}^{(t)})$ (from predicted frame encoder)
\Require $\hat{\mathbf{V}}_v^{(1)}$, $\mathbf{I}_v^{(1)}$, $\hat{\mathbf{V}}_{rv}^{(1)}$, $\mathbf{I}_{rv}^{(1)}$ (from reference frame decoder)
\Require $\hat{\mathbf{V}}_v^{(t-1)}$, $\hat{\mathbf{C}}_{rv}^{(t-1)}$ (from previous frame buffer)
\State // Geometry
\State $\mathbf{W}_v^{(1)}=RAHT(\hat{\mathbf{V}}_v^{(1)},J)$
\State $\widehat{\Delta\mathbf{V}}_v^{(t)}=IRAHT(\hat{\mathbf{V}}_v^{(1)},\widehat{\mathbf{T}\Delta\mathbf{V}}_v^{(t)},J)$
\State $\hat{\mathbf{V}}_v^{(t)}=\hat{\mathbf{V}}_v^{(t-1)}+\widehat{\Delta\mathbf{V}}_v^{(t)}$
\State $\hat{\mathbf{V}}^{(t)}=\hat{\mathbf{V}}_v^{(t)}(\mathbf{I}_v^{(1)})$
\State // Color
\State $\mathbf{W}_{rv}^{(1)}=RAHT(\hat{\mathbf{V}}_{rv}^{(1)},J)$
\State $\widehat{\Delta\mathbf{C}}_{rv}^{(t)}=IRAHT(\hat{\mathbf{V}}_{rv}^{(1)},\widehat{\mathbf{T}\Delta\mathbf{C}}_{rv}^{(t)},J)$
\State $\hat{\mathbf{C}}_{rv}^{(t)}=\hat{\mathbf{C}}_{rv}^{(t-1)}+\widehat{\Delta\mathbf{C}}_{rv}^{(t)}$
\State $\hat{\mathbf{C}}_r^{(t)}=\hat{\mathbf{C}}_{rv}^{(t)}(\mathbf{I}_{rv}^{(1)})$
\Ensure $\hat{\mathbf{V}}^{(t)}$, $\mathbf{F}^{(1)}$, $\hat{\mathbf{C}}_r^{(t)}$ (to renderer)
\Ensure $\hat{\mathbf{V}}_v^{(t)}$, $\hat{\mathbf{C}}_{rv}^{(t)}$ (to previous frame buffer)
\end{algorithmic}
\end{algorithm}

\subsubsection{Geometry encoding and decoding}

At the encoder, for frame $t$, as for frame $1$, the vertices $\mathbf{V}^{(1)}$, or equivalently the vertices $\hat{\mathbf{V}}^{(1)}$, are voxelized.  However, for frame $t>1$ the voxelization occurs with attributes $\mathbf{V}^{(t)}$.  In this sense, the vertices $\mathbf{V}^{(t)}$ are projected back to the reference frame, where they are voxelized like attributes.  As for frame $1$, this produces a possibly slightly shorter list $\hat{\mathbf{V}}_v^{(1)}$ along with a list of indices $\mathbf{I}_v^{(1)}$ such that $\hat{\mathbf{V}}^{(1)}=\hat{\mathbf{V}}_v^{(1)}(\mathbf{I}_v^{(1)})$.  In addition, it produces an equal-length list of representative attributes, $\mathbf{V}_v^{(t)}$.  Such a list is produced every frame.  Therefore the previous frame can be used as a prediction.  The prediction residual $\Delta\mathbf{V}_v^{(t)}=\mathbf{V}_v^{(t)}-\hat{\mathbf{V}}_v^{(t-1)}$ is transformed, quantized with stepsize $\Delta_{motion}$, inverse transformed, and added to the prediction to obtain the reproduction $\hat{\mathbf{V}}_v^{(t)}$, which goes into the frame buffer.  The quantized transform coefficients are entropy coded.  We use adaptive RLGR as the entropy coder.

At the decoder, the entropy code for the quantized transform coefficients of the prediction residual is received, entropy decoded, inverse transformed, inverse quantized, and added to the prediction to obtain $\hat{\mathbf{V}}_v^{(t)}$, which goes into the frame buffer.  Finally $\hat{\mathbf{V}}^{(t)}=\hat{\mathbf{V}}_v^{(t)}(\mathbf{I}_v^{(1)})$ is sent to the renderer.

\subsubsection{Color encoding and decoding}

At the encoder, for frame $t>1$, as for frame $t=1$, the refined vertices $\hat{\mathbf{V}}_r^{(1)}$, are voxelized with attributes $\mathbf{C}_r^{(t)}$.  In this sense, the colors $\mathbf{C}_r^{(t)}$ are projected back to the reference frame, where they are voxelized.  As for frame $t=1$, this produces a significantly shorter list $\hat{\mathbf{V}}_{rv}^{(1)}$ along with a list of indices $\mathbf{I}_{rv}^{(1)}$ such that $\hat{\mathbf{V}}_r^{(1)}=\hat{\mathbf{V}}_{rv}^{(1)}(\mathbf{I}_{rv}^{(1)})$.  In addition, it produces a list of representative attributes, $\mathbf{C}_{rv}^{(t)}$.  Such a list is produced every frame.  Therefore the previous frame can be used as a prediction.  The prediction residual $\Delta\mathbf{C}_{rv}^{(t)}=\mathbf{C}_{rv}^{(t)}-\hat{\mathbf{C}}_{rv}^{(t-1)}$ is transformed, quantized with stepsize $\Delta_{color,inter}$, inverse transformed, and added to the prediction to obtain the reproduction $\hat{\mathbf{C}}_{rv}^{(t)}$, which goes into the frame buffer.  The quantized transform coefficients are entropy coded.  We use adaptive RLGR as the entropy coder.

At the decoder, the entropy code for the quantized transform coefficients of the prediction residual is received, entropy decoded, inverse transformed, inverse quantized, and added to the prediction to obtain $\hat{\mathbf{C}}_{rv}^{(t)}$, which goes into the frame buffer.  Finally $\hat{\mathbf{C}}_r^{(t)}=\hat{\mathbf{C}}_{rv}^{(t)}(\mathbf{I}_{rv}^{(1)})$ is sent to the renderer.


\subsection{Rendering for visualization and distortion computation}
\label{ssec:rendering}
The decompressed  dynamic triangle  cloud $\lbrace\hat{\mathbf{V}}^{(t)},\hat{\mathbf{C}}_r^{(t)},\mathbf{F}^{(t)}\rbrace_{t=1}^N$ may have varying density across triangles resulting in some holes or transparent looking regions, which are not satisfactory for visualization.
We apply the triangle refinement function on the set of vertices and faces from Algorithm \ref{alg:refine} and produce the redundant representation 
$\lbrace\hat{\mathbf{V}}_r^{(t)},\hat{\mathbf{C}}_r^{(t)},\mathbf{F}_r^{(t)}\rbrace_{t=1}^N$. This sequence consists of a dynamic point cloud $\lbrace\hat{\mathbf{V}}_r^{(t)},\hat{\mathbf{C}}_r^{(t)}\rbrace_{t=1}^N$, whose colored points lie in the surfaces of triangles given by $\lbrace\hat{\mathbf{V}}_r^{(t)},\mathbf{F}_r^{(t)}\rbrace_{t=1}^N$. This representation is further refined using a similar method to increase the spatial resolution by adding a linear interpolation function for the color attributes as shown in Algorithm \ref{alg:refine_interpolate}.
The output is a denser point cloud, denoted by $\lbrace\hat{\mathbf{V}}_{rr}^{(t)},\hat{\mathbf{C}}_{rr}^{(t)}\rbrace_{t=1}^N$.  We use this denser point cloud for visualization and distortion computation in the experiments described in the next section.
\begin{algorithm}
\caption{Refinement and Color Interpolation}
\label{alg:refine_interpolate}
\begin{algorithmic}[1]
\Require ${\mathbf{V}}_r,{\mathbf{C}}_r$, $\mathbf{F}_r$, $U_{interp}$
\State $\mathbf{V}_i=\mathbf{V}_r(\mathbf{F}_r(:,i),:)$, $i=1,2,3$ // $i$th vertex of all faces
\State $\mathbf{C}_i=\mathbf{C}_r(\mathbf{F}_r(:,i),:)$, $i=1,2,3$ // color on $i$-th vertex
\State Initialize $\mathbf{V}_{rr}=\mathbf{C}_{rr}=$ empty list
\For{$i=0$ to $U_{interp}$}
	\For{$j=0$ to $U_{interp}-i$}
    	\State $\mathbf{V}_{rr}=[\mathbf{V}_{rr};\mathbf{V}_1+(\mathbf{V}_2-\mathbf{V}_1)i/U_{interp}+(\mathbf{V}_3-\mathbf{V}_1)j/U_{interp}]$
        \State $\mathbf{C}_{rr}=[\mathbf{C}_{rr};\mathbf{C}_1+(\mathbf{C}_2-\mathbf{C}_1)i/U_{interp}+(\mathbf{C}_3-\mathbf{C}_1)j/U_{interp}]$
    \EndFor
\EndFor
\Ensure $\mathbf{V}_{rr},\mathbf{C}_{rr}$
\end{algorithmic}
\end{algorithm}

\section{Experiments}
\label{sec:experiments}

In this section, we evaluate the RD performance of our system, for both intra-frame and inter-frame coding, for both color and geometry, under a variety of different error metrics.  Our baseline for comparison to previous work is intra-frame coding of colored voxels using octree coding for geometry \cite{meagher_octtree,Ochotta_2004,schnabel_2006,Kammerl2012} and RAHT coding for colors \cite{queiroz_raht}.

\subsection{Dataset}

\begin{figure*}
\begin{subfigure}[b]{0.33\textwidth}
\centering
\includegraphics[height=7cm]{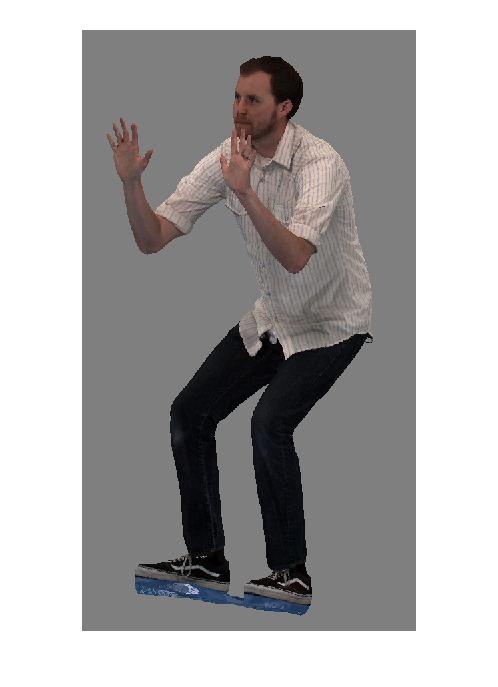}
\caption{Man }
\end{subfigure}
\begin{subfigure}[b]{0.33\textwidth}
\centering 
\includegraphics[height=7cm]{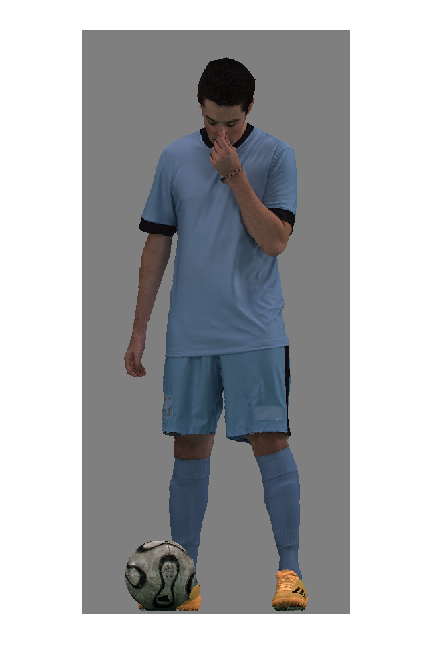}
\caption{Soccer}
\end{subfigure}
\begin{subfigure}[b]{0.33\textwidth}
\centering
\includegraphics[height=7cm]{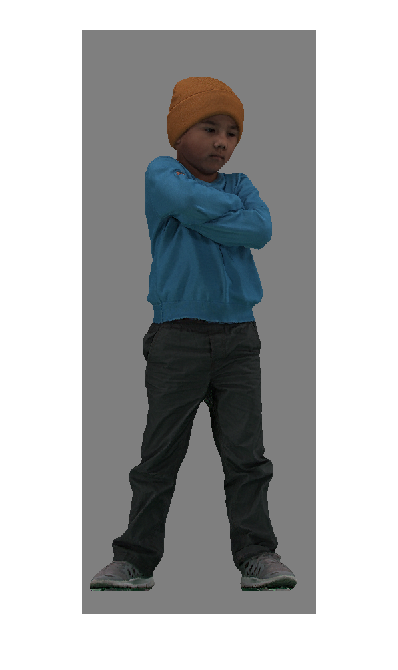}
\caption{Breakers}
\end{subfigure}
\caption{Initial frames of datasets {\em Man}, {\em Soccer}, and {\em Breakers}.}
\label{fig:datasets}
\end{figure*}

We use triangle cloud sequences derived from the Microsoft HoloLens Capture (HCap) mesh sequences \emph{Man}, \emph{Soccer}, and \emph{Breakers}\footnote{formally known as
{\em 2014\textunderscore 04\textunderscore 30\textunderscore Test\textunderscore 4ms},
{\em 2014\textunderscore 11\textunderscore 07\textunderscore Soccer\textunderscore Guy\textunderscore tra-ditional\textunderscore Take4},
and
{\em 2014\textunderscore 11\textunderscore 14\textunderscore Breakers\textunderscore modern\textunderscore minis\textunderscore Take4},
respectively}.  The initial frame from each sequence is shown in Figures~\ref{fig:datasets}a-c.
In the HCap sequences, each frame is a triangular mesh.  The frames are partitioned into groups of frames (GOFs).  Within each GOF, the meshes are consistent, i.e., the connectivity is fixed but the positions of the triangle vertices evolve in time.  We construct a triangle cloud from each mesh at time $t$ as follows.  For the vertex list $\mathbf{V}^{(t)}$ and face list $\mathbf{F}^{(t)}$, we use the vertex and face lists directly from the mesh.  For the color list $\mathbf{C}^{(t)}$, we upsample each face by factor $U=10$ to create a list of refined vertices, and then sample the mesh's texture map at the refined vertices.  The geometric data are scaled to fit in the unit cube $[0,1]^3$.  Our voxel size is $2^{-J}\times 2^{-J}\times 2^{-J}$, where $J=10$ is the maximum depth of the octree.  All sequences are $30$ frames per second.  The overall statistics are described in Table~\ref{tab:hcap_stats}.

\begin{table}
\begin{tabular}{|c|c|c|c|c|c|c|}\hline
Sequence & \# frm & \# GOF & $\vert \mathbf{V}\vert$/f & $\vert \mathbf{F}\vert$/f & voxels/f \\ \hline
Man &$200$&$7$&$11027$&$19978$&$561198$ \\ 
Soccer &$493$&$159$&$18187$&$33349$&$505803$ \\ 
Breakers &$496$&$156$&$12702$&$23178$&$411162$ \\
\hline
\end{tabular}
\caption{Dataset statistics.  Number of frames, number of GOFs (i.e., number of reference frames), and average number of vertices and faces per reference frame, in the original HCap datasets, and average number of occupied voxels per frame after voxelization with respect to reference frames.  All sequences are~30 fps.  For voxelization, all HCap meshes were upsampled by a factor of $U=10$, normalized to a $1\times1\times1$ bounding cube, and then voxelized into voxels of size $2^{-J}\times2^{-J}\times2^{-J}$, $J=10$.}
\label{tab:hcap_stats}
\end{table}


\subsection{Distortion metrics}

Comparing algorithms for compressing colored 3D geometry poses some challenges because there is no single agreed upon metric or distortion measure for this type of data.  Even if one attempts to separate the photometric and geometric aspects of distortion, there is often an interaction between the two.  We consider several metrics for both color and geometry to evaluate different aspects of our compression system.

\subsubsection{Projection distortion}

One common approach to evaluating the distortion of compressed colored geometry relative to an original is to render both the original and compressed versions of the colored geometry from a particular point of view, and compare the rendered images using a standard image distortion measure such as PSNR.


One question that arises with this approach is which viewpoint, or set of viewpoints, should be used.  Another question is which method of rendering should be used.  We choose to render from six viewpoints, by voxelizing the colored geometry of the refined and interpolated dynamic point cloud $\lbrace\hat{\mathbf{V}}_{rr}^{(t)},\hat{\mathbf{C}}_{rr}^{(t)}\rbrace_{t=1}^N$ described in Section~\ref{ssec:rendering},  and projecting the voxels onto the six faces of the bounding cube, using orthogonal projection.  For a cube of size $2^J\times2^J\times2^J$ voxels, the voxelized object is projected onto six images each of size $2^J\times2^J$ pixels.  If multiple occupied voxels project to the same pixel on a face, then the pixel takes the color of the occupied voxel closest to the face, i.e., hidden voxels are removed.  If no occupied voxels project to a pixel on a face, then the pixel takes a neutral gray color.  The mean squared error over the six faces and over the sequence is reported as PSNR separately for each color component: Y, U, and V.  We call this the {\em projection distortion}.

The projection distortion measures color distortion directly, but it also measures geometry distortion indirectly.  Thus we will report the projection distortion as a function of the motion stepsize ($\Delta_{motion}$) for a fixed color stepsize ($\Delta_{color}$), and vice versa, to understand the independent effects of geometry and color compression on this measure of quality.

\subsubsection{Matching distortion}

A {\em matching distortion} is a generalization of the Hausdorff distance commonly used to measure the difference between geometric objects \cite{MekuriaLTC16}.  Let $S$ and $T$ be source and target sets of points, and let $s\in S$ and $t\in T$ denote points in the sets, with color components (here, luminances) $Y(s)$ and $Y(t)$, respectively.  For each $s\in S$ let $t(s)$ be a point in $T$ {\em matched} (or {\em assigned}) to $s$, and likewise for each $t\in T$ let $s(t)$ be a point in $S$ assigned to $t$.  The functions $t(\cdot)$ and $s(\cdot)$ need not be invertible.  Commonly used functions are the nearest neighbor assignments
\begin{eqnarray}
t^*(s) & = & \arg\min_{t\in T} d^2(s,t) \\
s^*(t) & = & \arg\min_{s\in S} d^2(s,t)
\end{eqnarray}
where $d^2(s,t)$ is a geometric distortion measure such as the squared error $d^2(s,t)=||s-t||_2^2$.  Given matching functions $t(\cdot)$ and $s(\cdot)$, the {\em forward} (one-way) {\em mean squared matching distortion} has geometric and color components
\begin{eqnarray}
d_G^2(S\rightarrow T) & = & \frac{1}{|S|} \sum_{s\in S} ||s-t(s)||_2^2 \label{eqn:fwdmatdst_G} \\
d_Y^2(S\rightarrow T) & = & \frac{1}{|S|} \sum_{s\in S} |Y(s)-Y(t(s))|_2^2
\end{eqnarray}
while the {\em backward mean squared matching distortion} has geometric and color components
\begin{eqnarray}
d_G^2(S\leftarrow T) & = & \frac{1}{|T|} \sum_{t\in T} ||t-s(t)||_2^2 \\
d_Y^2(S\leftarrow T) & = & \frac{1}{|T|} \sum_{t\in T} |Y(t)-Y(s(t))|_2^2 \label{eqn:bwdmatdst_C}
\end{eqnarray}
and the {\em symmetric mean squared matching distortion} has geometric and color components
\begin{eqnarray}
d_G^2(S,T) & = & \max\{d_G^2(S\rightarrow T),d_G^2(S\leftarrow T)\} \label{eqn:symmatdst_G} \\
d_Y^2(S,T) & = & \max\{d_Y^2(S\rightarrow T),d_Y^2(S\leftarrow T)\}. \label{eqn:symmatdst_C}
\end{eqnarray}
In the event that the sets $S$ and $T$ are not finite, the averages in (\ref{eqn:fwdmatdst_G})-(\ref{eqn:bwdmatdst_C}) can be replaced by integrals, e.g., $\int_S ||s-t(s)||_2^2 d\mu(s)$ for an appropriate measure $\mu$ on $S$.

The forward, backward, and symmetric {\em Hausdorff} matching distortions are similarly defined, with the averages replaced by maxima (or integrals replaced by suprema).

Though there can be many variants on these measures, for example using averages in (\ref{eqn:symmatdst_G})-(\ref{eqn:symmatdst_C}) instead of maxima, or using other norms or robust measures in (\ref{eqn:fwdmatdst_G})-(\ref{eqn:bwdmatdst_C}), these definitions are consistent with those in \cite{MekuriaLTC16} when $t^*(\cdot)$ and $s^*(\cdot)$ are used as the matching functions.
(Though we do not use them here, matching functions other than $t^*(\cdot)$ and $s^*(\cdot)$, which take color into account and are smoothed, such as in \cite{QueirozC16-simple}, may yield distortion measures that are better correlated with subjective distortion.)
In this paper, for consistency with the literature, we use the symmetric mean squared matching distortion with matching functions $t^*(\cdot)$ and $s^*(\cdot)$.

For each frame $t$, we compute the matching distortion between sets $S^{(t)}$ and $T^{(t)}$, which are obtained by the sampling  the texture map of the original HCap data to obtain a high resolution point  cloud  $(\mathbf{V}_{rr}^{(t)},\mathbf{C}_{rr}^{(t)})$ with $J=10$ and $U=40$. We compare its colors and vertices  to the decompressed and color interpolated high resolution point cloud $(\mathbf{\hat{V}}_{rr}^{(t)},\mathbf{\hat{C}}_{rr}^{(t)})$ with interpolation factor $U_{interp}=4$ described in Section~\ref{ssec:rendering}\footnote{Note that the original triangle cloud was obtained by sampling the HCap data with upsampling factor $U=10$.  Thus by interpolating the decompressed triangle cloud with $U_{interp}=4$, the overall number of vertices and triangles is the same as obtained by sampling the original HCap data with upsampling factor $U=40$.}.   We then voxelize both point clouds and  compute the mean squared matching distortion over all frames as
\begin{eqnarray}
\bar d_G^2 & = & \frac{1}{N}\sum_{t=1}^N d_G^2(S^{(t)},T^{(t)}) \\
\bar d_Y^2 & = & \frac{1}{N}\sum_{t=1}^N d_Y^2(S^{(t)},T^{(t)})
\end{eqnarray}
and we report the geometry and color components of the matching distortion in dB as
\begin{eqnarray}
PSNR_G & = & -10\log_{10}\frac{\bar d_G^2}{3W^2} \\
PSNR_Y & = & -10\log_{10}\frac{\bar d_Y^2}{255^2} \\
\end{eqnarray}
where $W=1$ is the width of the bounding cube.

Note that even though the geometry and color components of the distortion measure are separate, there is an interaction: The geometry affects the matching, and hence affects the color distortion.  Thus we will report the color component of the matching distortion as a function of the color stepsize ($\Delta_{color}$) for a fixed motion stepsize ($\Delta_{motion}$), and vice versa, to understand the independent effects of geometry and color compression on color quality.  We report the geometry component of the matching distortion as a function only of the motion stepsize ($\Delta_{motion}$), since color compression does not affect the geometry under the assumed matching functions $t^*(\cdot)$ and $s^*(\cdot)$.

\subsubsection{Triangle cloud distortion}

In our setting, the input and output of our system are the triangle clouds $(\mathbf{V}^{(t)},\mathbf{F}^{(t)},\mathbf{C}^{(t)})$ and $(\hat{\mathbf{V}}^{(t)},\mathbf{F}^{(t)},\hat{\mathbf{C}}^{(t)})$. Thus natural measures of  distortion for our system are 
\begin{eqnarray}
PSNR_G = -10\log_{10}\left( \frac{1}{N} \sum_{t=1}^{N} \frac{||\mathbf{V}_r^{(t)}-\hat{\mathbf{V}}_r^{(t)}||^2_2}{3W^2N_r^{(t)}}\right) \label{eqn:triclddst_G0} \\
PSNR_Y = -10\log_{10}\left( \frac{1}{N} \sum_{t=1}^{N} \frac{||\mathbf{Y}_r^{(t)} -\hat{\mathbf{Y}}_r^{(t)}||^2_2}{255^2 N_r^{(t)}}\right) ,
\label{eqn:triclddst_C0}
\end{eqnarray}
where $\mathbf{Y}_r^{(t)}$ is the first (i.e, luminence) column of the $N_r^{(t)}\times3$ matrix of color attributes $\mathbf{C}_r^{(t)}$ and $W=1$ is the width of the bounding cube.  These represent the average geometric and luminance distortions across the faces of the triangles.  $PSNR_U$ and $PSNR_V$ can be similarly defined.

However for rendering we use higher resolution versions of the triangles, in which both the vertices and the colors are interpolated up from $\mathbf{V}_r^{(t)}$ and $\mathbf{C}_r^{(t)}$ using Algorithm~\ref{alg:refine_interpolate} to obtain higher resolution vertices and colors $\mathbf{V}_{rr}^{(t)}$ and $\mathbf{C}_{rr}^{(t)}$. We use the following distortion measures as very close approximations of (\ref{eqn:triclddst_G0}) and (\ref{eqn:triclddst_C0}): 
\begin{eqnarray}
PSNR_G = -10\log_{10}\left( \frac{1}{N} \sum_{t=1}^{N} \frac{||\mathbf{V}_{rr}^{(t)}-\hat{\mathbf{V}}_{rr}^{(t)}||^2_2}{3W^2N_{rr}^{(t)}}\right) \label{eqn:triclddst_G} \\
PSNR_Y = -10\log_{10}\left( \frac{1}{N} \sum_{t=1}^{N} \frac{||\mathbf{Y}_{rr}^{(t)} -\hat{\mathbf{Y}}_{rr}^{(t)}||^2_2}{255^2 N_{rr}^{(t)}}\right) ,
\label{eqn:triclddst_C}
\end{eqnarray}
where $\mathbf{Y}_{rr}^{(t)}$ is the first (i.e, luminence) column of the $N_{rr}^{(t)}\times3$ matrix of color attributes $\mathbf{C}_{rr}^{(t)}$ and $W=1$ is the width of the bounding cube.  $PSNR_U$ and $PSNR_V$ can be similarly defined.

\subsubsection{Transform coding distortion}

For the purposes of rate-distortion optimization, and other rapid distortion computations, it is more convenient to use an internal distortion measure: the distortion between the input and output of the tranform coder.  We call this the {\em transform coding distortion}, defined in dB as
\begin{eqnarray}
PSNR_G = -10\log_{10} \left( \frac{1}{N} \sum_{t=1}^{N}\frac{||\mathbf{V}_v^{(t)} -\hat{\mathbf{V}}_v^{(t)}||^2_2}{3W^2 N_v^{(t)}}\right) \label{eqn:trncoddst_G} \\
PSNR_Y = -10\log_{10} \left( \frac{1}{N} \sum_{t=1}^{N}\frac{||\mathbf{Y}_{rv}^{(t)} -\hat{\mathbf{Y}}_{rv}^{(t)}||^2_2}{255^2 N_{rv}^{(t)}}\right) ,
\label{eqn:trncoddst_C}
\end{eqnarray}
where $\mathbf{Y}_{rv}^{(t)}$ is the first (i.e, luminence) column of the $N_{rv}^{(t)}\times3$ matrix $\mathbf{C}_{rv}^{(t)}$.  $PSNR_U$ and $PSNR_V$ can be similarly defined.  Unlike (\ref{eqn:triclddst_G})-(\ref{eqn:triclddst_C}), which are based on system inputs and outputs $\mathbf{V}^{(t)},\mathbf{C}^{(t)}$ and $\hat{\mathbf{V}}^{(t)},\hat{\mathbf{C}}^{(t)}$, (\ref{eqn:trncoddst_G})-(\ref{eqn:trncoddst_C}) are based on the voxelized quantities $\mathbf{V}_v^{(t)},\mathbf{C}_{rv}^{(t)}$ and $\hat{\mathbf{V}}_v^{(t)},\hat{\mathbf{C}}_{rv}^{(t)}$, which are defined for reference frames in Algorithm~\ref{alg:Iencoder} (Steps~3, 6, and 9) and for predicted frames in Algorithm~\ref{alg:Pencoder} (Steps~2, 7, 9, and~14).  The squared errors in the two cases are essentially the same, but are weighted differently: one by face and one by voxel.
\subsection{Rate metrics}
As with the distortion, we  report bit rates for compression of a whole sequence, for geometry and color. We compute the bit rate averaged over a sequence, in megabits per second, as
\begin{align}
R_{Mbps}=\frac{bits}{1024^2 N }30\;\text{[Mbps]}
\end{align}
where $N$ is the number of frames in the sequence, and $bits$ is the total number of bits used to encode the color or geometry information of the sequence.
Also, we report the bit rate in bits per voxel as
\begin{align}
R_{bpv}=\frac{bits}{\sum_{t=1}^{N}N_{rv}^{(t)}} \;\text{[bpv]}
\end{align}
where $N_{rv}^{(t)}$ is the number of occupied voxels in frame $t$ and again $bits$ is the total number of bits used to encode color or geometry for the whole sequence. The number of voxels of a given frame $N_{rv}^{(t)}$ depends on the voxelization used. For example in our triangle cloud encoder, within a GOF all frames have the same number of voxels, because the voxelization of attributes is done with respect to the reference frame. For our triangle encoder in all intra mode, each frame will have a different number of voxels.  
\subsection{Intra-frame coding}

We first examine intra-frame coding of triangle clouds, and compare it to intra-frame coding of voxelized point clouds.  To obtain the voxelized point clouds, we voxelize the original mesh-based sequences {\em Man}, {\em Soccer}, and {\em Breakers} by refining each face in the original sequence by upsampling factor $U=10$, and voxelizing to level $J=10$.  For each sequence, and each frame $t$, this produces a list of occupied voxels $\mathbf{V}_{rv}^{(t)}$ and their colors $\mathbf{C}_{rv}$.

\subsubsection{Intra-frame coding of geometry}

We compare our method for coding geometry in reference frames with the previous state-of-the-art for coding geometry in single frames.  The previous state-of-the art for coding the geometry of voxelized point clouds \cite{meagher_octtree,Ochotta_2004,schnabel_2006,Kammerl2012} codes the set of occupied voxels $\mathbf{V}_{rv}^{(t)}$ by entropy coding the octree description of the set.  In contrast, our method first approximates the set of occupied voxels by a set of triangles, and then codes the triangles as a triple $(\mathbf{V}_v^{(t)},\mathbf{F}^{(t)},\mathbf{I}_v^{(t)})$.  The vertices $\mathbf{V}_v^{(t)}$ are coded using octrees plus \emph{gzip}, the faces $\mathbf{F}^{(t)}$ are coded directly with \emph{gzip}, and the indices $\mathbf{I}_v^{(t)}$ are coded using run-length encoding plus \emph{gzip} as described in Section~\ref{subsec:reference_geometry}.  When the geometry is smooth, relatively few triangles need to be used to approximate it.  In such cases, our method gains because the list of vertices $\mathbf{V}_v^{(t)}$ is much shorter than the list of occupied voxels $\mathbf{V}_{rv}^{(t)}$, even though the list of triangle indices $\mathbf{F}^{(t)}$ and the list of repeated indices $\mathbf{I}_v^{(t)}$ must also be coded.

Taking all bits into account, Table~\ref{tab:intra_frame_geometry} shows the bit rates for both methods in megabits per second (Mbps) and bits per occupied voxel (bpv) averaged over the sequences.  Our method reduces the bit rate needed for intra-frame coding of geometry by a factor of 5-10, breaking through the 2.5 bpv rule-of-thumb for octree coding.

\begin{table}
\centering
\begin{tabular}{|c|c|c|c|c|}\hline
& \multicolumn{2}{|c|}{Previous} & \multicolumn{2}{|c|}{Ours} \\ \hline
Sequence & Mbps & bpv & Mbps & bpv \\ \hline
Man & $50.7$ &$3.20$ & $5.24$ &$0.33$ \\ 
Soccer & $37.6$ & $2.61$& $6.39$ &$0.44$ \\ 
Breakers & $43.7$ &$3.28$ & $4.88$ &$0.36$ \\
\hline
\end{tabular}
\caption{Intra-frame coding of the geometry of voxelized point clouds.  ``Previous'' refers to our implementation of the octree coding approach described in \cite{meagher_octtree,Ochotta_2004,schnabel_2006,Kammerl2012}.}
\label{tab:intra_frame_geometry}
\end{table}

While it is true that approximating the geometry by triangles is generally not lossless, in this case the process is lossless because our ground truth datasets are already described in terms of triangles.

\subsubsection{Intra-frame coding of color}

Our method of coding color in reference frames is identical with the state-of-the art for coding color in single frames, using transform coding based on RAHT, described in \cite{queiroz_raht}.  For reference, the rate-distortion results for color intra-frame coding are shown in Figure~\ref{fig:plot_color_RD_hybrid_intra} (where we compare to color inter-frame coding).

\subsection{Inter/intra-frame coding: transform coding distortion rate curves}

We next examine hybrid inter-frame plus intra-frame coding (here called inter/intra-frame coding) of triangle clouds using the transform coding distortion, and compare it to intra-frame only coding of triangle clouds.  We show that temporal prediction provides substantial gains for geometry across all sequences, and significant gains for color on one of the three sequences.

\subsubsection{Inter/intra-frame coding of geometry}

\begin{figure}[tb]
\begin{center}
\begin{minipage}[b]{\linewidth}
  \centering
  \centerline{\includegraphics[width=\linewidth]{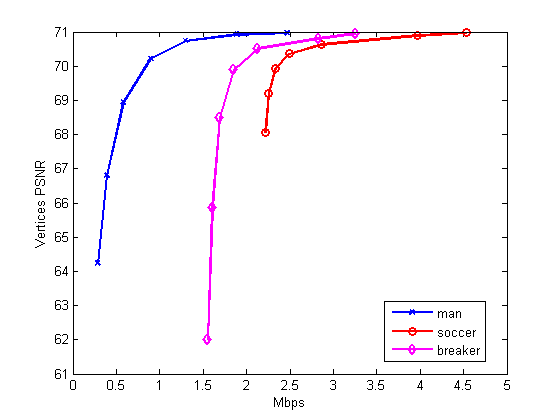}}
  \centerline{(a) RD curves for motion compression.}\medskip
\end{minipage}
\end{center}
\begin{center}
\begin{minipage}[b]{\linewidth}
  \centering
  \centerline{\includegraphics[width=\linewidth]{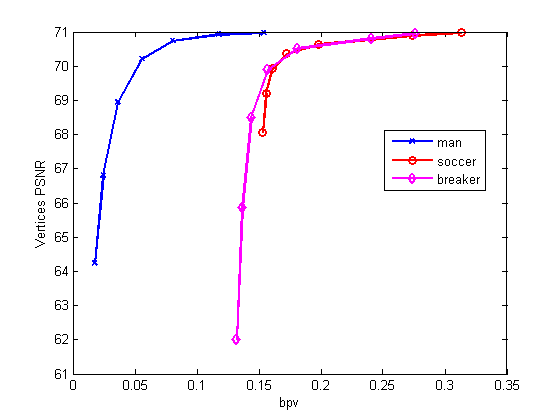}}
  \centerline{(b) RD curves for motion compression.}\medskip
\end{minipage}
\end{center}
\caption{RD curves for geometry compression.  Rates include all geometry information.}
\label{fig:RD_motion1}
\end{figure}

\begin{figure}[tb]
\begin{center}
\begin{minipage}[b]{1.00\linewidth}
  \centering
  \centerline{\includegraphics[width=\linewidth]{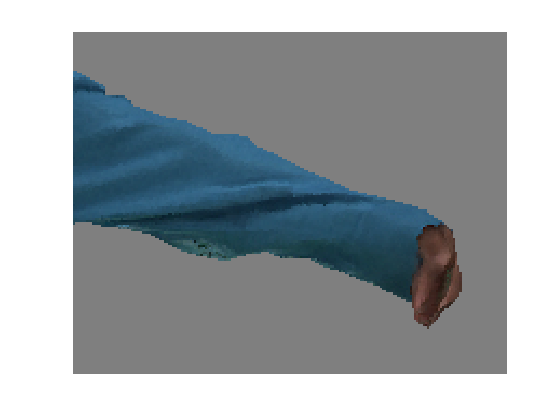}}
  \centerline{(a) original}\medskip
\end{minipage}

\begin{minipage}[b]{1.00\linewidth}
  \centering
  \centerline{\includegraphics[width=\linewidth]{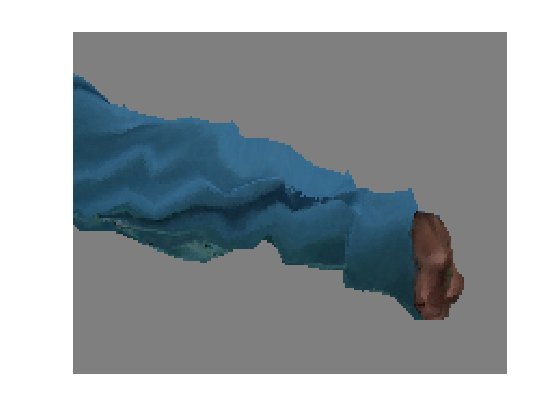}}
  \centerline{(b) 62 dB (1.6 Mbps for all geometry information)}\medskip
\end{minipage}

\begin{minipage}[b]{1.00\linewidth}
  \centering
  \centerline{\includegraphics[width=\linewidth]{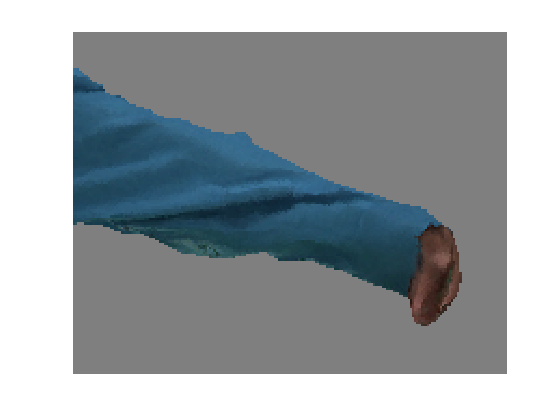}}
  \centerline{(c) 70.5 dB (2.2 Mbps for all geometry information)}\medskip
\end{minipage}
\end{center}
%
\caption{Visual quality of geometry compression.}
\label{fig:RD_motion2}
\vspace{-0.4cm}
\end{figure}

Figure~\ref{fig:RD_motion1} shows the geometry transform coding distortion $PSNR_G$ (\ref{eqn:trncoddst_G}) as a function of the bit rate needed for geometry information in inter/intra-frame coding of the sequences {\em Man}, {\em Soccer}, and {\em Breakers}.  It can be seen that the geometry PSNR saturates, at relatively low bit rates, at the highest fidelity possible for a given voxel size $2^{-J}$, which is 71 dB for $J=10$.  In Figure~\ref{fig:RD_motion2} we show on the \emph{Breakers} sequence that quality within 0.5 dB of this limit appears to be sufficiently close to that of the original voxelization without quantization.  At this quality, for \emph{Man}, \emph{Soccer}, and \emph{Breakers} sequences, the encoder in inter/intra (hybrid) mode has geometry bit rates of about $1.2$, $2.7$, and $2.2$ Mbps (0.07, 0.19, 0.17 bpv), respectively.  For comparison, the encoder in all-intra mode has geometry bit rates of $5.24$, $6.39$, and $4.88$ Mbps (0.33, 0.44, 0.36 bpv),  respectively, as shown in Table~\ref{tab:intra_frame_geometry}.  Thus the intra-inter mode has a geometry bit rate savings of a factor of $2$-$5$ over our intra-frame coding only, and a factor of 13-45 over previous intra-frame octree coding.

%

\begin{figure*}
\begin{center}
\begin{subfigure}[b]{0.33\textwidth}
\includegraphics[width=\textwidth]{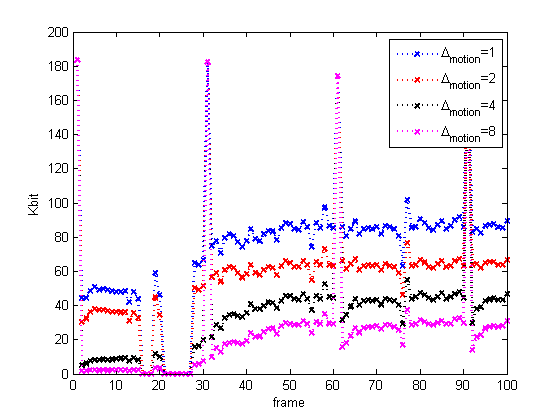}
\caption{Man }
\end{subfigure}
\begin{subfigure}[b]{0.33\textwidth}
\includegraphics[width=\textwidth]{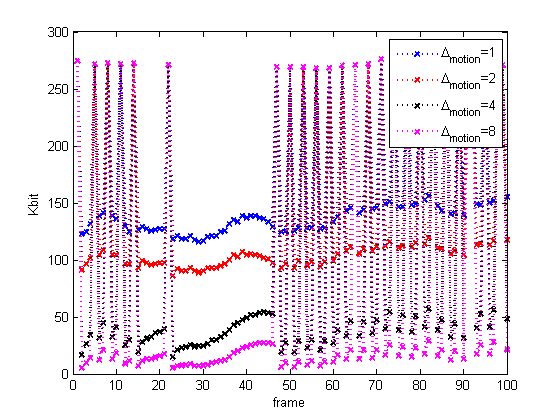}
\caption{Soccer}
\end{subfigure}
\begin{subfigure}[b]{0.33\textwidth}
\includegraphics[width=\textwidth]{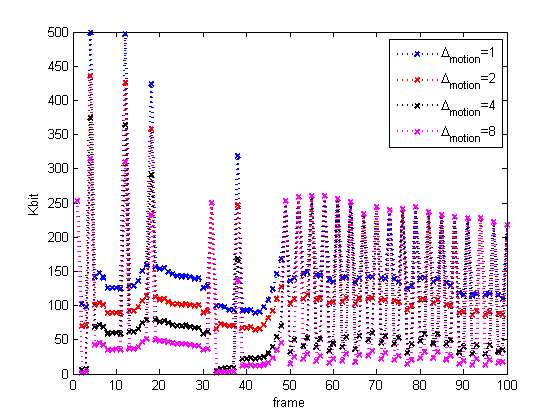}
\caption{Breaker}
\end{subfigure}
\end{center}
\caption{Kilobits/frame required to code the geometry information for each frame for different values of the motion residual quantization stepsize  $\Delta_{motion} \in \lbrace 1,2,4,8\rbrace$.  Reference frames encode $\mathbf{V}^{(1)}_v$ using octree coding plus {\em gzip} and encode $\mathbf{I}_v^{(1)}$ using run-length coding plus {\em gzip}. Predicted frames encode their motion residuals $\Delta \mathbf{V}^{(t)}$ using transform coding.}
\label{fig:motion_frame1}
\end{figure*}

\begin{figure*}
\begin{center}
\begin{subfigure}[b]{0.33\textwidth}
\includegraphics[width=\textwidth]{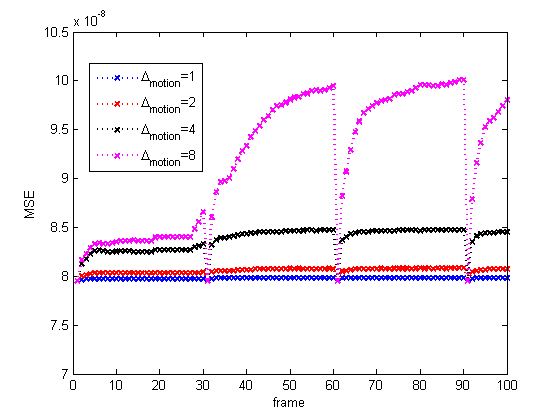}
\caption{Man}
\end{subfigure}
\begin{subfigure}[b]{0.33\textwidth}
\includegraphics[width=\textwidth]{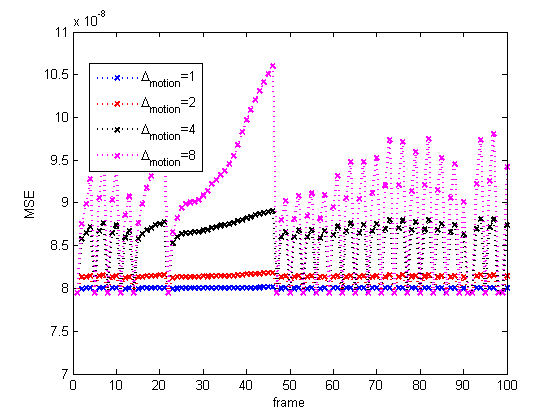}
\caption{Soccer}
\end{subfigure}
\begin{subfigure}[b]{0.33\textwidth}
\includegraphics[width=\textwidth]{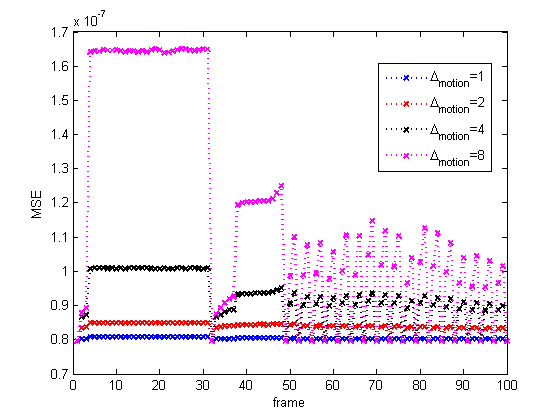}
\caption{Breaker}
\end{subfigure}
\end{center}
\caption{Mean squared quantization error required to code the geometry information for each frame for different values of the motion residual quantization stepsize  $\Delta_{motion} \in \lbrace 1,2,4,8\rbrace$. Reference frames encode $\mathbf{V}^{(1)}_v$ using octrees; hence the distortion is due to quantization error is $\epsilon^2$. Predicted frames encode their motion residuals $\Delta \mathbf{V}^{(t)}$ using transform coding.}
\label{fig:motion_frame2}
\end{figure*}

A temporal analysis is provided in Figures~\ref{fig:motion_frame1} and~\ref{fig:motion_frame2}.  Figure~\ref{fig:motion_frame1} shows the number of kilobits per frame needed to encode the geometry information for each frame. The number of bits for the reference frames are dominated by their octree descriptions, while the number of bits for the predicted frames depends on the quantization stepsize for motion residuals, $\Delta_{motion}$. We observe that a significant bit reduction can be achieved by lossy coding of residuals.  For $\Delta_{motion}=4$, there is more than a 3x reduction in bit rate for inter-frame coding relative to intra-frame coding.

Figure~\ref{fig:motion_frame2} shows the mean squared quantization error
\begin{equation}
MSE_G = \frac{1}{N} \sum_{t=1}^{N} \frac{||\mathbf{V}_v^{(t)}-\hat{\mathbf{V}}_v^{(t)}||^2_2}{3W^2N_v^{(t)}},
\end{equation}
which corresponds to the $PSNR_G$ in (\ref{eqn:trncoddst_G}).
Note that for reference frames, the mean squared error is well approximated by
\begin{equation}
\frac{||\mathbf{V}_v^{(1)}-\hat{\mathbf{V}}_v^{(1)}||^2_2}{3W^2N_v^{(t)}} \approx \frac{2^{-2J}}{12} \stackrel{\Delta}{=} \epsilon^2.
\end{equation}
Thus for reference frames, the $MSE_G$ falls to $\epsilon^2$, while for predicted frames, the $MSE_G$ rises from $\epsilon^2$ depending on the motion stepsize $\Delta_{motion}$.

\subsubsection{Intra/inter-frame coding of color}

To evaluate color coding, first we consider separate quantization stepsizes for reference and predicted frames
$\Delta_{color,intra}$ and $\Delta_{color,inter}$ respectively. Both take values in $\{1, 2, 4, 8,$ $16, 32, 64\}$.

\begin{figure}
  \centering
  {\includegraphics[width=\linewidth]{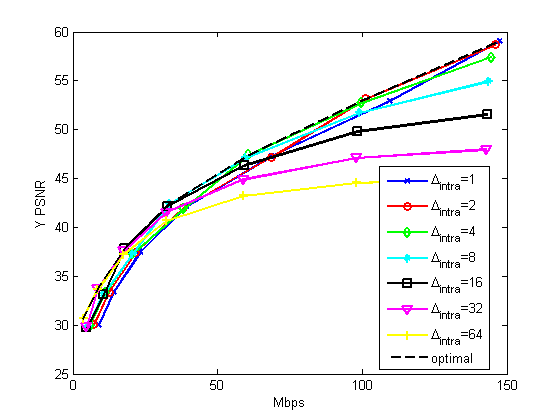}}
  {\includegraphics[width=\linewidth]{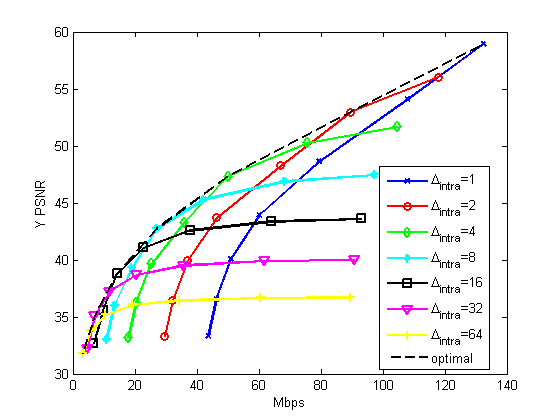}}
 {\includegraphics[width=\linewidth]{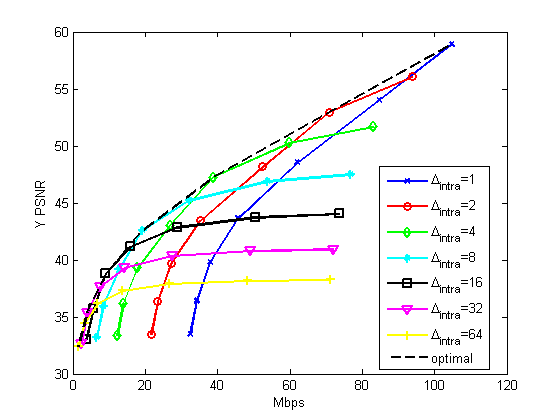}}
  \caption{Luminance (Y) component rate-distortion performances of (top) \emph{Man}, (middle) \emph{Soccer} and (bottom) \emph{Breakers} sequences, for different intra-frame stepsizes $\Delta_{color,intra}$. Rate includes all ($Y$, $U$, $V$) color information.}
  \label{fig:RD_color}
\end{figure}

\begin{figure}
  \centering
  {\includegraphics[width=\linewidth]{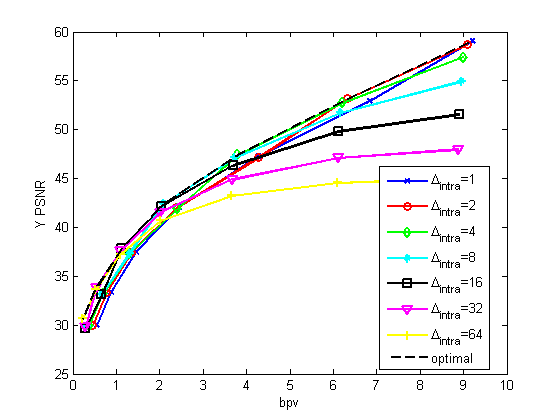}}
  {\includegraphics[width=\linewidth]{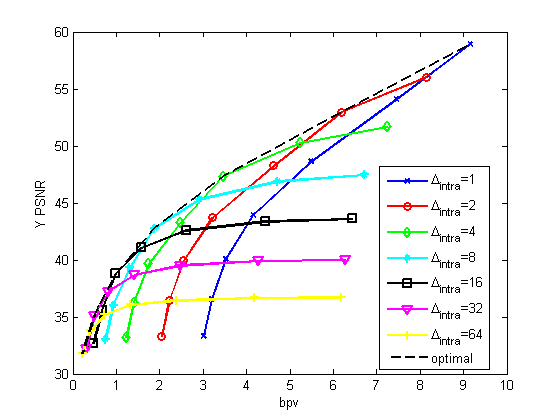}}
 {\includegraphics[width=\linewidth]{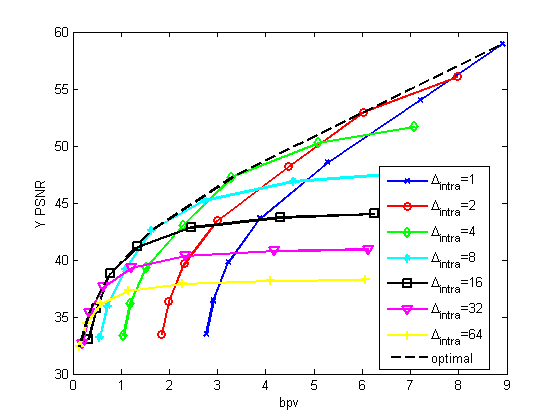}}
  \caption{Luminance (Y) component rate-distortion performances of (top) \emph{Man}, (middle) \emph{Soccer} and (bottom) \emph{Breakers} sequences, for different intra-frame stepsizes $\Delta_{color,intra}$. Same as Figure \ref{fig:RD_color} but rate is in bits-per-voxel.}
  \label{fig:RD_color_bpv}
\end{figure}

Figures~\ref{fig:RD_color} and \ref{fig:RD_color_bpv} shows the color transform coding distortion $PSNR_Y$ (\ref{eqn:trncoddst_C}) as a function of the bit rate 
(Mbps and bpv respectively) needed for all ($Y$, $U$, $V$) color information for inter/intra-frame coding of the sequences {\em Man}, {\em Soccer}, and {\em Breakers}, for different combinations of $\Delta_{color,intra}$ and $\Delta_{color,inter}$, where each colored curve corresponds to a fixed value of $\Delta_{color,intra}$.  It can be seen that the optimal RD curve is obtained by choosing $\Delta_{color,intra}=\Delta_{color,inter}$, as shown in the dashed line.

Next, we consider equal quantization stepsizes for reference and predicted frames, hereafter designated simply $\Delta_{color}$.

\begin{figure}
\centering
 {\includegraphics[width=\linewidth]{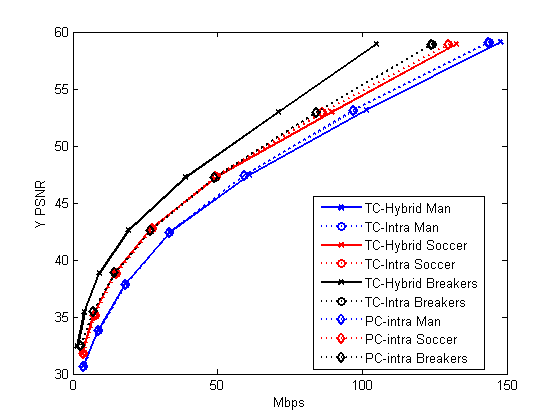}}
  {\includegraphics[width=\linewidth]{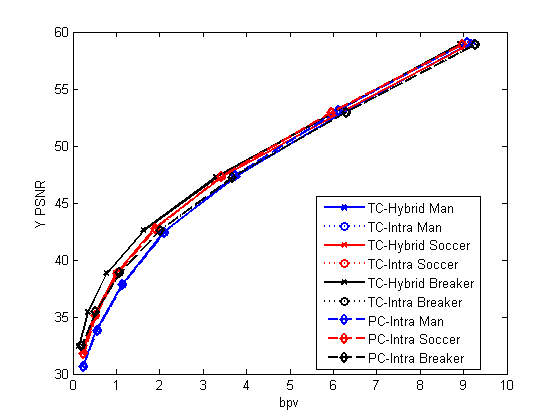}}

\caption{Inter/intra-frame coding vs.\ intra-frame only coding.  The bit rate contains all ($Y$, $U$, $V$) color information, although the distortion is only the luminance ($Y$) PSNR.
}
\label{fig:plot_color_RD_hybrid_intra}
\end{figure}

Figure~\ref{fig:plot_color_RD_hybrid_intra} shows the color transform coding distortion $PSNR_Y$ (\ref{eqn:trncoddst_C}) as a function of the bit rate needed for all ($Y$, $U$, $V$) color information for inter/intra-frame coding and intra-frame only coding
on the sequences {\em Man}, {\em Soccer}, and {\em Breakers}.  We observe that inter/intra-frame coding outperforms intra-frame only coding by $2$-$3$ dB for the \emph{Breakers} sequence. However, for the \emph{Man} and \emph{Soccer} sequences, their RD performances are similar.  Further investigation is needed on when and how gains can be achieved by predictive coding of color.

\begin{figure*}
\begin{center}
\begin{subfigure}[b]{0.33\textwidth}
\includegraphics[width=\textwidth]{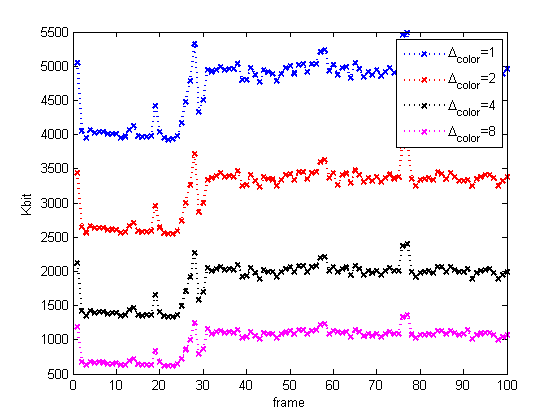}
\caption{Man}
\end{subfigure}
\begin{subfigure}[b]{0.33\textwidth}
\includegraphics[width=\textwidth]{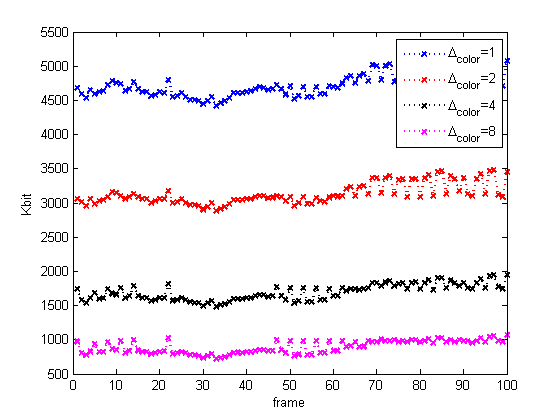}
\caption{Soccer}
\end{subfigure}
\begin{subfigure}[b]{0.33\textwidth}
\includegraphics[width=\textwidth]{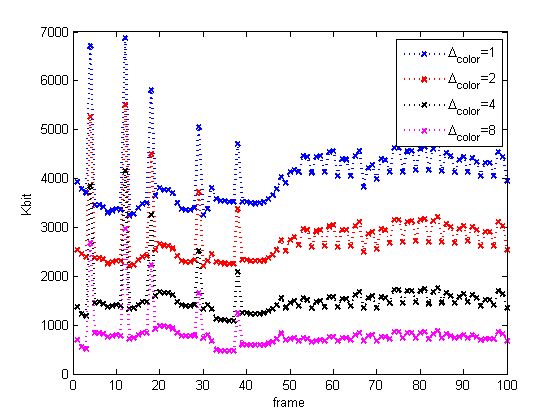}
\caption{Breaker}
\end{subfigure}
\end{center}
\caption{Kilobits/frame required to code the color information for each frame for different values of the color residual quantization stepsize  $\Delta_{color} \in \lbrace 1,2,4,8\rbrace$.  Reference frames encode their colors $\mathbf{C}_{rv}^{(1)}$ and predicted frames encode their color residuals $\Delta \mathbf{C}_{rv}^{(t)}$ using transform coding.}
\label{fig:color_frame_rate}
\end{figure*}

\begin{figure*}
\begin{center}
\begin{subfigure}[b]{0.33\textwidth}
\includegraphics[width=\textwidth]{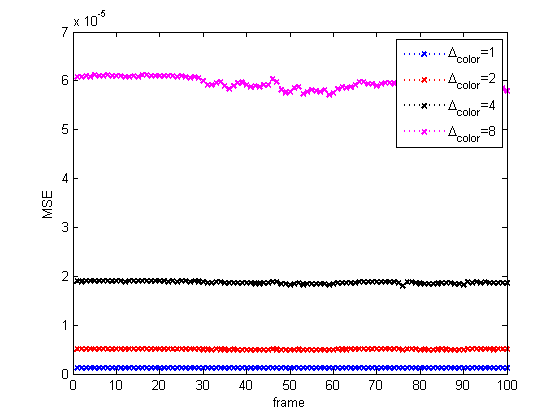}
\caption{Man}
\end{subfigure}
\begin{subfigure}[b]{0.33\textwidth}
\includegraphics[width=\textwidth]{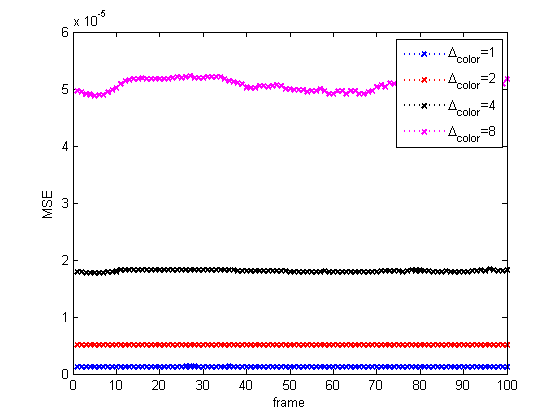}
\caption{Soccer}
\end{subfigure}
\begin{subfigure}[b]{0.33\textwidth}
\includegraphics[width=\textwidth]{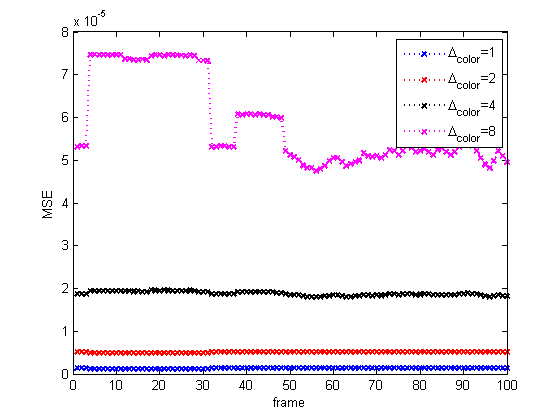}
\caption{Breaker}
\end{subfigure}
\end{center}
\caption{Mean squared quantization error required to code the color information for each frame for different values of the color residual quantization stepsize  $\Delta_{color} \in \lbrace 1,2,4,8\rbrace$. Reference frames encode their colors $\mathbf{C}_{rv}^{(1)}$ and predicted frames encode their color residuals $\Delta \mathbf{C}_{rv}^{(t)}$ using transform coding.}
\label{fig:color_frame_distortion}
\end{figure*}

A temporal analysis is provided in Figures~\ref{fig:color_frame_rate} and~\ref{fig:color_frame_distortion}. In Figure~\ref{fig:color_frame_rate} we show the bit rates (Kbit) to compress the color information for the first $100$ frames of all sequences. We observe that, as expected, for smaller values of $\Delta_{color}$ the bit rates are higher, for all frames. For \emph{Man} and \emph{Soccer} sequences we observe that the bit rates do not vary much  from reference frames to predicted frames; however in the \emph{Breakers} sequence, it is clear that for all values of $\Delta_{color}$ the reference frames have much higher bit rates compared to predicted frames, which confirms the results from Figure~\ref{fig:plot_color_RD_hybrid_intra}, where inter/intra-frame coding provides gains with respect to intra-frame only coding of triangle clouds for the \emph{Breakers} sequence, but not for the {\em Man} and {\em Soccer} sequences. 
In Figure~\ref{fig:color_frame_distortion} we show the MSE of the Y color component for the first $100$ frames of all sequences. For  $\Delta_{color} \leq 4$ the error is uniform across all frames and sequences.

\subsubsection{Comparison to dynamic mesh compression}

We now compare our results to the dynamic mesh compression in \cite{anis_2016}, which uses a distortion measure similar to the transform coding distortion measure, and reports results on a version of the {\em Man} sequence.

For geometry coding, Figure~5 in \cite{anis_2016} shows that when their geometry distortion is 70.5 dB, their geometry bit rate is about 0.45 bpv. As shown in Figure~\ref{fig:RD_motion1}, at the same distortion, our bit rate is about 0.07 bpv, which is lower than their bit rate by a factor of 6x or more.

For color coding, Figure~5 in \cite{anis_2016} shows that when their color distortion is 40 dB, their color bit rate is about 0.8 bpv.  As shown in Figure~\ref{fig:plot_color_RD_hybrid_intra}, at the same distortion, our bit rate is about 1.8 bpv.

Overall, their bit rate would be about $0.45+0.8=1.3$ bpv, while our bit rate would be about $0.07+1.8=1.9$ bpv.  However it should be cautioned that the sequence compressed in \cite{anis_2016} is not the original {\em Man} sequence used in our work but rather a smooth mesh fit to a low-resolution voxelization ($J=9$) of the sequence.  Hence it has smoother color as well as smoother geometry, and should be easier to code.  Nevertheless, it is a point of comparison.

\subsection{Inter/intra-frame coding: triangle cloud, projection, and matching distortion-rate curves}

In this section we show distortion rate curves using the triangle cloud, projection, and matching distortion measures. All distortions in this section are computed from high resolution triangle clouds generated from the original HCap data, and from the decompressed triangle clouds.
For computational complexity reasons, we show results only for the \emph{Man} sequence, and consider only its first four GOFs (120 frames).

\subsubsection{Geometry coding}

\begin{figure}
\begin{center}
\begin{subfigure}[b]{0.48\textwidth}
\includegraphics[width=\textwidth]{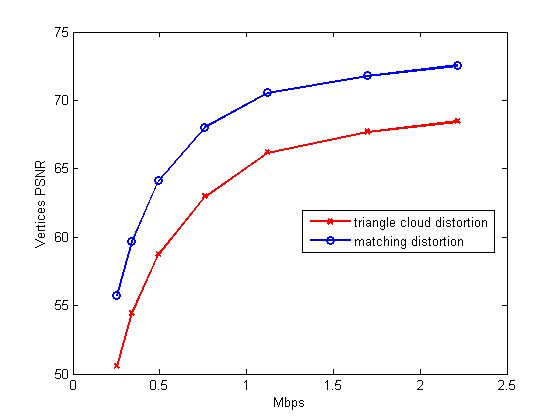}
\caption{Geometry distortion vs geometry bit rate [Mbps]}
\end{subfigure}
\begin{subfigure}[b]{0.48\textwidth}
\includegraphics[width=\textwidth]{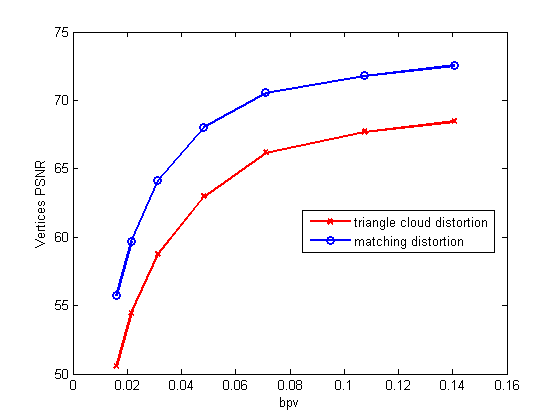}
\caption{Geometry distortion vs geometry bit rate [bpv]}
\end{subfigure}
\end{center}
\caption{RD  curves for geometry triangle cloud and matching distortion vs.\ geometry bit rates.}
\label{fig:motion_D1D3}
\end{figure}

First we analyze the triangle cloud distortion and matching distortion of geometry as a function of geometry bit rate.  The RD plots are shown in Figure~\ref{fig:motion_D1D3}. We observe that both distortion measures start saturating at the same point as for the transform coding distortion: around  $\Delta_{motion}=4$.  However for these distortion measures the saturation is not as pronounced.
This suggest that these distortion measures are quite sensitive to small amounts of geometric distortion.

\begin{figure}
\begin{center}
\begin{subfigure}[b]{0.48\textwidth}
\includegraphics[width=\textwidth]{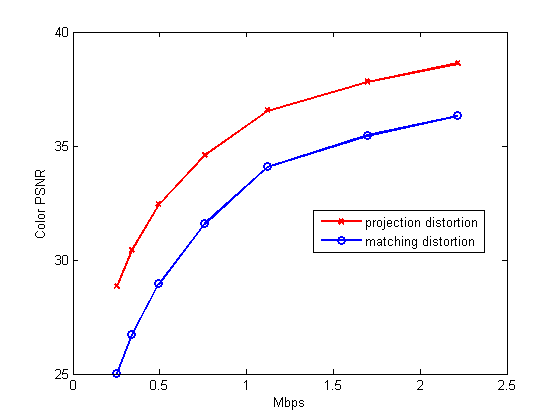}
\caption{Color distortion vs geometry bit rate [Mbps]}
\end{subfigure}
\begin{subfigure}[b]{0.48\textwidth}
\includegraphics[width=\textwidth]{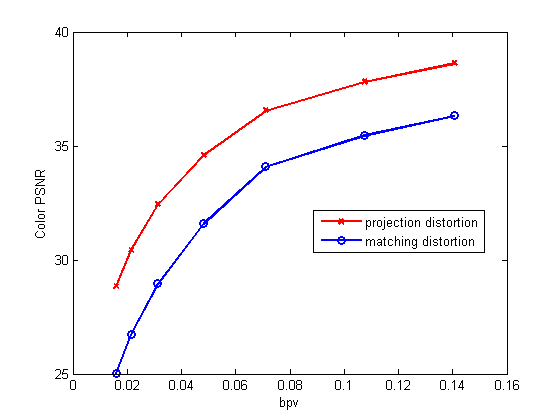}
\caption{Color distortion vs geometry bit rate [bpv]}
\end{subfigure}
\end{center}
\caption{RD  curves for color triangle cloud and matching distortion vs.\ geometry bit rates.  The color stepsize is set to $\Delta_{color}=1$.}
\label{fig:motion_D2D3}
\end{figure}

Next we study the effect of geometry compression on color quality.  In Figure~\ref{fig:motion_D2D3} we show the $Y$ component PSNR for the projection and matching distortion measures.
The color has been compressed at the highest bit rate considered, using the same quantization step for intra and inter color coding, $\Delta_{color}=1$.  Surprisingly, we observe a significant influence of the geometry compression on these color distortion measures, particular for $\Delta_{motion}>4$.  This indicates very high sensitivity to geometric distortion of the projection distortion measure and the color component of the matching distortion measure.  This hyper-sensitivity can be explained as follows.  For the projection distortion measure, geometric distortion causes local shifts of the image.  As is well known, PSNR, as well as other image distortion measures including SSIM, fall apart upon image shifts.  For the matching metric, since the matching functions $s^*$ and $t^*$ depend only on geometry, geometric distortion causes inappropriate matches, which affect the color distortion across those matches.


\subsubsection{Color coding}

\begin{figure}
\begin{center}
\begin{subfigure}[b]{0.48\textwidth}
\includegraphics[width=\textwidth]{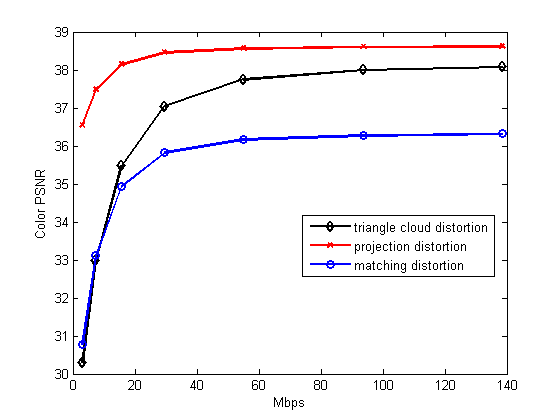}
\caption{Color distortion vs color bit rate [Mbps]}
\end{subfigure}
\begin{subfigure}[b]{0.48\textwidth}
\includegraphics[width=\textwidth]{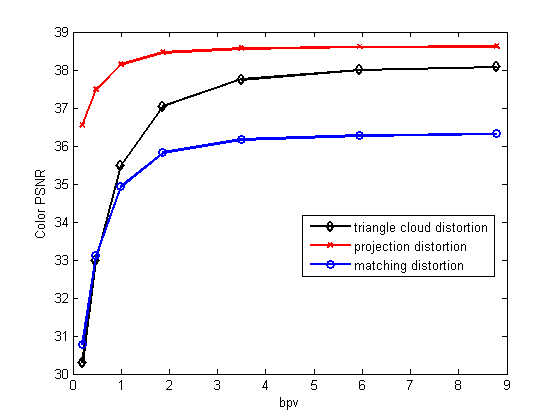}
\caption{Color distortion vs color bit rate [bpv]}
\end{subfigure}
\end{center}
\caption{RD curves for color triangle cloud, projection, and  matching distortion vs color bit rates.  The motion stepsize is set to $\Delta_{motion}=1$.}
\label{fig:color_D1D2D3}
\end{figure}

Finally we analyze the RD curve for color coding as a function of color bit rate. We plot $Y$ component PSNR for the triangle cloud, projection, and matching distortion measures in Figure~\ref{fig:color_D1D2D3}.  For this experiment we consider the color quantization steps equal for intra and inter coded frames.  The motion step is set to $\Delta_{motion}=1$.  For all three distortion measures, the PSNR saturates very quickly.  Apparently, this is because the geometry quality severely limits the color quality under any of the these three distortion measures, even when the geometry quality is high ($\Delta_{motion}=1$).  In particular, when $\Delta_{motion}=1$, for color quantization stepsizes smaller than $\Delta_{color}=8$, color quality does not improve significantly under these distortion measures, while under the transform coding distortion measure, the PSNR continues to improve, as shown in Figures~\ref{fig:RD_color} and~\ref{fig:RD_color_bpv}.  Whether the hyper-sensitivity of the color projection and color  matching distortion measures to geometric distortion are perceptually justified is questionable, but open to further investigation.

\subsubsection{Comparison to dynamic point cloud compression}

Notwithstanding possible issues with the color projection distortion measure, it provides an opportunity to compare our results on dynamic triangle cloud compression to the results on dynamic point cloud compression in \cite{QueirozC16-simple}.  Like us, \cite{QueirozC16-simple} reports results on a version of the {\em Man} sequence, using the projection distortion measure.

Figure~\ref{fig:RD_color_bpv} shows that for triangle cloud compression, the projection distortion reaches 38.5 dB at around 2 bpv.  In comparison, Figure~10a in \cite{QueirozC16-simple} shows that for dynamic point cloud compression, the projection distortion reaches 38.5 dB at around 3 bpv.  Hence it seems that our dynamic triangle cloud compression may be more efficient than point cloud compression under the projection distortion measure.  However it should be cautioned that the sequence compressed in \cite{QueirozC16-simple} is a lower resolution ($J=9$) version of the {\em Man} sequence rather than the higher resolution version ($J=10$) used in our work.  Moreover, Figure~\ref{fig:RD_color_bpv} in our paper reports the distortion between the original signal (with uncoded color and uncoded geometry) to the coded signal (with coded color and coded geometry), while Figure~10a in \cite{QueirozC16-simple} reports the distortion between the signal with uncoded color and coded geometry to the signal with coded color and identically coded geometry.  In the latter case, the saturation of the color measure due to geometric coding is not apparent.

\section{Conclusion}
\label{sec:discussion}

When coding for video, the representation of the input to the encoder and the representation of the output of the decoder are clear: sequences of rectangular arrays of pixels.  Furthermore, distortion measures between the two representations are well accepted in practice.

In contrast, when coding for augmented reality, as of yet, the representation of the input to the encoder and the representation of the output of the decoder are not yet widely agreed upon in the research community.  This is because the there are many types of sensing scenarios and rigs, each requiring a different process for fusing raw camera data into the encoder input representation.  Likewise, there are many varieties of display scenarios and devices, each requiring a different process for rendering the decoder output representation.  Naturally, distortion measures between any such representations are also not yet widely agreed upon.

Two leading candidates for the codec's representation for augmented reality to this point have been dynamic meshes and dynamic point clouds.  Each has its advantages and disadvantages.  Dynamic meshes fit well into the traditional graphic pipeline and have high compression efficiency.  However, they do not accommodate well the noise and non-surface topologies typically present in real time live capture.  Conversely, dynamic point clouds are well-suited for representing noise and non-surface topologies, but are difficult to interpolate in space and time, making them difficult to compress by exploiting spatial and temporal redundancies.

In this paper, we proposed dynamic polygon clouds, which have the advantages of both meshes and point clouds, without their disadvantages.  We provided detailed algorithms on how to compress them, and we used a variety of distortion measures to evaluate their performance.

For intra-frame coding of geometry, we showed that compared to the previous state-of-the-art for intra-frame coding of the geometry of voxelized point clouds, our method reduces the bit rate by a factor of 5-10 with negligible (but non-zero) distortion, breaking through the 2.5 bpv rule-of-thumb for lossless coding of geometry in voxelized point clouds.  Intuitively, these gains are achieved by reducing the representation from a dense list of points to a less dense list of vertices and faces.

For inter-frame coding of geometry, we showed that compared to our method of intra-frame coding of geometry, we can reduce the bit rate by a factor of 3 or more.  For inter/intra-frame (hybrid) coding, this results in a geometry bit rate savings of a factor of 2-5 over intra-frame coding only.  Intuitively, these gains are achieved by coding the motion prediction residuals.  Multiplied by the 5-10 x improvement of our intra-frame coding compared to previous octree-based intra-frame coding, we have demonstrated a 13-45 x reduction in bit rate over previous octree-based intra-frame coding.

For inter-frame coding of color, we showed that compared to our method of intra-frame coding of color (which is the same as the current state-of-the-art for intra-frame coding of color \cite{queiroz_raht}), our method reduces the bit rate by about 30\% or alternatively increases the PSNR by about 2 dB (at the relevant level of quality) for one of our three sequences.  For the other two sequences, we found little improvement in performance relative to intra-frame coding of color.  This is a matter for further investigation, but one hypothesis is that the gain is dependent upon the quality of the motion estimation.  Intuitively,  gains are achieved by coding the color prediction residuals, and the color prediction is accurate only if the motion estimation is accurate.

We compared our results on triangle cloud compression to recent results in dynamic mesh compression and dyanmic point cloud compression.  The comparisons are imperfect due to somewhat different datasets and distortion measures, which likely favor the earlier work.  However, they indicate that compared to dynamic mesh compression, our geometry coding may have a bit rate 6x lower, while our color coding may have a bit rate 2.25x higher.  At the same time, compared to dynamic point cloud compression, our overall bit rate may be about 33\% lower.

Our work also revealed the hyper-sensitivity of distortion measures such as the color projection and color matching distortion measures to geometry coding.

Future work includes better transforms and better entropy coders, RD optimization, better motion compensation, and more perceptually relevant distortion measures and post-processing filtering.

\section{Acknowledgment}

The authors would like to thank the Microsoft HoloLens Capture (HCap) team for making their data available to this research, and would also like to thank the Microsoft Research Interactive 3D (I3D) team for many discussions.

%



\vfill\pagebreak

\bibliographystyle{IEEEbib}
\bibliography{references}

\end{document}